\documentclass[journal]{IEEEtran}
\usepackage{amsmath,amsfonts}
\usepackage{algorithmic}
\usepackage{algorithm}
\usepackage{array}
\usepackage[caption=false,font=normalsize,labelfont=sf,textfont=sf]{subfig}
\usepackage{textcomp}
\usepackage{stfloats}
\usepackage{url}
\usepackage{verbatim}
\usepackage{graphicx}
\usepackage{cite}
\usepackage{microtype}
\usepackage{amsmath}
\usepackage{pifont}
\usepackage{booktabs}
\usepackage{multirow}
\usepackage{framed}
\usepackage{color}
\usepackage{url}
\usepackage{hyperref}
\usepackage[table,xcdraw]{xcolor}
\allowdisplaybreaks[4] 

\begin{document}

\title{Risk Assessment and Vulnerability Identification of Energy-Transportation Infrastructure Systems to Extreme Weather}

\author{Jiawei~Wang,~\IEEEmembership{Graduate Student Member,~IEEE,}
        Qinglai~Guo,~\IEEEmembership{Fellow,~IEEE,}
        Haotian~Zhao,~\IEEEmembership{Member,~IEEE,}
		Bin~Wang,~\IEEEmembership{Member,~IEEE,}
        and~Hongbin~Sun,~\IEEEmembership{Fellow,~IEEE}
\thanks{J. Wang, Q. Guo, H. Zhao, B. Wang, and H. Sun are with the Department of Electrical Engineering, Tsinghua University, Beijing 100084, China. (\textit{Corresponding Author: Qinglai Guo, Haotian Zhao. Email: guoqinglai@tsinghua.edu.cn, zhaohaotian@mail.tsinghua.edu.cn})}
\thanks{This work was supported by the National Key R\&D Program of China (No. 2022YFB2404000) and the National Natural Science Foundation of China (NSFC) under Grant (No. 52321004).}
}

\markboth{Journal of \LaTeX\ Class Files,~Vol.~14, No.~8, August~2021}%
{Shell \MakeLowercase{\textit{et al.}}: A Sample Article Using IEEEtran.cls for IEEE Journals}

\IEEEpubid{0000--0000/00\$00.00~\copyright~2021 IEEE}

\maketitle

\begin{abstract}
	The interaction between extreme weather events and interdependent critical infrastructure systems involves complex spatiotemporal dynamics. Multi-type emergency decisions within energy-transportation infrastructures significantly influence system performance throughout the extreme weather process. A comprehensive assessment of these factors faces challenges in model complexity, heterogeneous differences between energy and transportation systems, and cross-sector privacy. This paper proposes a risk assessment framework that integrates the heterogeneous energy and transportation systems in the form of a unified network flow model, which enables full accommodation of multiple types of energy-transportation emergency decisions while capturing the compound spatiotemporal impacts of extreme weather on both systems simultaneously. Based on this framework, a targeted method for identifying system vulnerabilities is further developed. This method employs neural network surrogates to achieve privacy protection and accelerated identification while maintaining consideration of system interdependencies. Numerical experiments demonstrate that the proposed framework and method can reveal the risk levels faced by urban infrastructure systems, identify vulnerabilities that should be prioritized for reinforcement, and strike a balance between accuracy and speed.
\end{abstract}


\section*{Nomenclature}
\addcontentsline{toc}{section}{Nomenclature}
{\linespread{0.98}\selectfont
\small
\subsection{Abbreviations}
\begin{IEEEdescription}[\IEEEusemathlabelsep\IEEEsetlabelwidth{AMoD}]
\item[EV] Electric vehicle
\item[LHS] Latin hypercube sampling
\item[SOC] State of charge
\item[CHP] Combined heating and power
\item[PV] Photovoltaic
\item[CV] Coefficient of variation
\item[MAPE] Mean absolute percentage error
\end{IEEEdescription}

\subsection{Parameters}
\begin{IEEEdescription}[\IEEEusemathlabelsep\IEEEsetlabelwidth{$x^{\text{drop}}_{\text{ori},\text{des},t}$}]
\item[$\Delta t$] The length of each time period
\item[$\sigma_i^{\text{power}}$] The economic value coefficient of power node $i$
\item[$\sigma_i^{\text{heat}}$] The economic value coefficient of heat node $i$

\item[$n_{\text{loc/time/soc}}$] Location/time/state of charge (SOC) attributes of a node in the time-space-energy augmented network of the traffic network
\item[$i_{\text{ori},\sim}, i_{\text{des},\sim}$] The start and end nodes' attributes of edge $i$ in the time-space-energy augmented network of the traffic network
\item[$i_{\text{dist}}$] The spatial distance of edge $i$ in the time-space-energy augmented network of the traffic network
\item[$i_{\text{delay}}$] The additional time delay number corresponding to edge $i$ in the time-space-energy augmented network of the traffic network
\item[$r_{i}^{\text{in},(\text{0})}$] The fixed economic value of traffic demand $i$
\item[$r_{i}^{\text{in},(\text{1})}$] The unit distance economic value of traffic demand $i$
\item[$r_{i}^{\text{time}}$] The unit time delay cost for traffic demand $i$
\item[$o_{\text{ori,des},t}$] Number of traffic demands requesting to move from origin ``ori'' to destination ``des'' in time period $t$
\item[$P^{\text{cha/dis}}$] Charging/discharging power of a single vehicle
\item[$C^{\text{cha/dis,max}}_f$] Maximum charging/discharging capacities at charging node $f$
\item[$c^{\text{0}}$] Initial SOC of each vehicle, also enforced as the required SOC at the final time period
\item[$t^{\text{max}}$] The final time period within the simulation range

\item[$u_{a,b,t}$] State of road or branch $(a,b)$ in time period $t$, where 0 indicates interruption and 1 indicates no interruption
\item[$v_{a,b,t}$] Performance level of road $(a,b)$ in time period $t$, i.e., the ratio of the average travel speed to the ideal state speed
\item[$t_{i,a}$] Time period when the flow on edge $i$ in the time-space-energy augmented network reaches location $a$
\item[$T_{a,b}$] Required travel time for road $(a,b)$ under normal conditions

\item[$\eta^{\text{uti}}_s$] The energy utilization efficiency of generator $s$
\item[$R^{\text{need}}_{i,j}$] Required repair resource threshold for power branch $(i,j)$
\item[$t_{i,j}^{\text{break}}$] The time when power line $(i,j)$ is destroyed by the disaster
\item[$t^{\text{repair}}_{i,j}$] The repair time required for branch $(i,j)$
\item[$\mu_i$] The efficiency of electric heating at node $i$
\end{IEEEdescription}

\IEEEpubidadjcol

\subsection{Sets}
\begin{IEEEdescription}[\IEEEusemathlabelsep\IEEEsetlabelwidth{$x^{\text{drop}}_{\text{ori},\text{des},t}$}]
\item[$\mathcal{T}$] The set of simulation time periods
\item[$\mathcal{N}$] The set of nodes in the power network
\item[$\mathcal{B}$] The set of branches in the power network
\item[$\mathcal{H}$] The set of nodes in the heat network
\item[$\mathcal{P}$] The set of pipes in the heat network
\item[$\mathcal{H}^{\text{load}}$] The set of heat load nodes in the heat network
\item[$\mathcal{E}$] The set of nodes in the time-space-energy augmented network of the traffic network
\item[$\mathcal{I}^{\text{type}}$] The set of edges of type ``type'' in the time-space-energy augmented network of the traffic network
\item[$\mathcal{G}$] The set of locations in the traffic network
\item[$\mathcal{F}$] The set of charging stations
\item[$\mathcal{S'}$] The set of nodes requiring fuel delivery
\item[$\text{path}(i)$] The set of roads chosen for edge $i$ in the time-space-energy augmented network
\end{IEEEdescription}

\subsection{Variables}
\begin{IEEEdescription}[\IEEEusemathlabelsep\IEEEsetlabelwidth{$x^{\text{drop}}_{\text{ori},\text{des},t}$}]
\item[$P^{(l*)}_{i,t}$] The pure active power load at power node $i$ at time period $t$
\item[$P^{(l)}_{i,t}$] The total load of power node $i$ at time period $t$
\item[$P^{(g)}_{i,t}$] Power output of the generator at node $i$ at time period $t$
\item[$H_{i,t}^{(l)}$] The heat load at heat node $i$ at time period $t$
\item[$H^{(g)}_{i,t}$] The heat power of the heat source at node $i$ at time period $t$

\item[$s_n, e_n$] The external injection and outflow at node $n$ in the time-space-energy augmented network of the traffic network
\item[$x^{\text{drop}}_{\text{ori},\text{des},t}$] Traffic demands that are not satisfied from origin ``ori'' to destination ``des'' in time period $t$
\item[$P_{f,t}^{\text{cha/dis}}$] Charging/discharging power at charging node $f$ in time period $t$
\item[$\xi^{\text{sup}}_{s,t}$] Fuel supply provided to fuel demand node $s$ in time period $t$
\item[$\theta_{g,t}$] Repair resources actively engaged in emergency repair work at location $g$ in time period $t$
\item[$x_i^{\text{type}}$] Flow of type ``type'' on edge $i$ in the time-space-energy augmented network of the traffic network

\item[$\xi_{s,t}$] Fuel reserve of node $s$ at time perisod $t$
\item[$s_{i,j,t}$] Indicator of whether branch $(i,j)$ is closed or disconnected at time period $t$
\item[$\delta_{i,j,t}$] Indicator of whether repair resources for branch $(i,j)$ exceed the threshold at time period $t$
\end{IEEEdescription}
}

\section{Introduction}

\IEEEPARstart{I}n recent years, climate change has led to more frequent and widespread occurrences of extreme weather events such as hurricanes and snowstorms \cite{JUFRI_SOTA_resilience_2019}. The destructive impact of extreme weather is neither static nor isolated but follows a certain spatiotemporal evolution trajectory, affecting multiple coupled systems simultaneously, particularly energy and transportation systems \cite{Sun_EnergyInternet_2024}. For example, Hurricane Irma (2017) caused power outages in Florida, with fallen trees and debris blocking transportation, leading to road closures and delaying power system recovery. Hurricane Harvey (2017) in Texas caused widespread power outages in Houston and triggered cascading failures in transportation, water, and communication systems, delaying relief efforts. In Zhengzhou, China (2021), a severe rainstorm flooded power lines and substations, causing blackouts, while water accumulation paralyzed transportation and hindered recovery efforts. These cases highlight the complexity of infrastructure interactions under extreme weather.

Regarding the resilience of energy systems under extreme weather, substantial research has considered emergency decision factors, incorporating elements such as microgrid formation, distributed energy resources, electric vehicles (EVs) or mobile energy storage, repair crews, preventive reinforcement, and fuel transportation. These studies analyze the risks faced by urban infrastructure systems during extreme weather and propose valuable action strategies \cite{Sturmer_Hardening_Lines_2024, Wang_Coordination_Restoration_2024, TAO2023_DAD, WANG2022_SeparableMESS, LIBODA2022_LTT,  WangKe_Heat_Reconfiguration_2023, QIU_Hierarchical_MARL_Repair_Crews_2023,Wang_Microgrid_Mobile_Sources_2024}, as summarized in Table \ref{table-literature-review-compare}. Among these studies, \cite{Sturmer_Hardening_Lines_2024} focuses only on pre-disaster preventive measures. \cite{Wang_Coordination_Restoration_2024, WANG2022_SeparableMESS, WangKe_Heat_Reconfiguration_2023, QIU_Hierarchical_MARL_Repair_Crews_2023, Wang_Microgrid_Mobile_Sources_2024} focus only on post-disaster restoration, optimizing system actions under the condition that all failures and damages have already occurred and are treated as fixed facts. Because these models consider only a single stage, they may face limitations when used to assess the overall risk of extreme weather events. \cite{TAO2023_DAD, LIBODA2022_LTT} incorporate multiple resilience-enhancing strategies that couple prevention and response, yet their extensive discrete-variable modeling and robust or stochastic optimization formulations make them overly complex and less scalable for flexibly accommodating diverse emergency decisions.

Meanwhile, although existing studies have made significant contributions by exploring various combinations of emergency decisions, most of them focus only on power systems, overlooking how interactions among different infrastructure systems, especially transportation systems, could influence assessments under extreme weather conditions. This could hinder an accurate understanding of risk levels. In extreme weather events, multiple types of emergency decisions rely on the actual mobility behaviors occurring within the transportation system, not only as an adjunct to the energy system. The transportation system, as an entity with its own behavioral characteristics, may also be affected by compounded impacts of extreme weather on its behavior and state. Therefore, if the analysis focuses only on the energy system and treats mobile resources only as boundary conditions moving between its nodes, it cannot fully capture the interdependencies among systems or the compound impacts of extreme weather on energy-transportation systems.

\begin{table*}[!t]
	\centering
	\renewcommand\arraystretch{0.95}
	\caption{Comparison of this paper's features with those of existing research.}
	\begin{tabular}{lcllcl}
	\hline
	\textbf{Ref.} &
	\multicolumn{1}{l}{\textbf{\begin{tabular}[c]{@{}l@{}}Cross-System \\ Analysis\end{tabular}}} &
	\textbf{\begin{tabular}[c]{@{}l@{}}Emergency \\ Actions\end{tabular}} &
	\textbf{\begin{tabular}[c]{@{}l@{}}Risk \\ Treatment\end{tabular}} &
	\multicolumn{1}{l}{\textbf{\begin{tabular}[c]{@{}l@{}}Privacy \\ Consideration\end{tabular}}} &
	\textbf{Approach} \\ \hline
	\rowcolor[HTML]{EFEFEF} 
	\cite{Sturmer_Hardening_Lines_2024} &
	\ding{55} &
	Line Hardening &
	Probability of Line Failure &
	\ding{55} &
	Monte Carlo \\
	\cite{Wang_Coordination_Restoration_2024} &
	\ding{55} &
	\begin{tabular}[c]{@{}l@{}}Electric Vehicles, \\ Repair Crew\end{tabular} &
	(Post-Disaster) &
	\ding{55} &
	Optimization \\
	\rowcolor[HTML]{EFEFEF} 
	\cite{TAO2023_DAD} &
	\ding{55} &
	\begin{tabular}[c]{@{}l@{}}Line Hardening, \\ Mobile Power Source, \\ Repair Crew\end{tabular} &
	Uncertainty Set of Ice Storms &
	\ding{55} &
	Robust Optimization \\
	\cite{WANG2022_SeparableMESS} &
	\ding{55} &
	\begin{tabular}[c]{@{}l@{}}Topology Reconfiguration, \\ Mobile Power Source, \\ Fuel Transport\end{tabular} &
	(Post-Disaster) &
	\ding{55} &
	Optimization \\
	\rowcolor[HTML]{EFEFEF} 
	\cite{LIBODA2022_LTT} &
	\checkmark &
	\begin{tabular}[c]{@{}l@{}}Topology Reconfiguration, \\ Fuel Transport\end{tabular} &
	Probability of Line Failure &
	\ding{55} &
	Stochastic Programming \\
	\cite{WangKe_Heat_Reconfiguration_2023} &
	\checkmark &
	\begin{tabular}[c]{@{}l@{}}Topology Reconfiguration of \\ Power and Heat Networks\end{tabular} &
	(Post-Disaster) &
	\ding{55} &
	Optimization \\
	\rowcolor[HTML]{EFEFEF} 
	\cite{QIU_Hierarchical_MARL_Repair_Crews_2023} &
	\checkmark &
	Repair Crew &
	(Post-Disaster) &
	\checkmark &
	Reinforcement Learning \\
	\cite{Wang_Microgrid_Mobile_Sources_2024} &
	\ding{55} &
	\begin{tabular}[c]{@{}l@{}}Topology Reconfiguration, \\ Mobile Power Source, \\ Repair Crew\end{tabular} &
	(Post-Disaster) &
	\checkmark &
	Reinforcement Learning \\ \hline
	\rowcolor[HTML]{EFEFEF} 
	\textbf{\begin{tabular}[c]{@{}l@{}}This\\ Paper\end{tabular}} &
	\textbf{\checkmark} &
	\textbf{\begin{tabular}[c]{@{}l@{}}Multi-Type \\ Emergency Decisions and \\ Flexible Combinations\end{tabular}} &
	\textbf{\begin{tabular}[c]{@{}l@{}}Compound Effects of Extreme \\ Weather on Energy-Transportation \\ System Across the Entire Process\end{tabular}} &
	\cellcolor[HTML]{EFEFEF}\textbf{\checkmark} &
	\textbf{\begin{tabular}[c]{@{}l@{}}Monte Carlo and\\ Surrogate-based \\ Optimizaion\end{tabular}} \\ \hline
	\end{tabular}
	\label{table-literature-review-compare}
\end{table*}

Urban infrastructure systems may be coupled at various levels, such as physical, geographical, informational, and logical. The spatiotemporal dynamic development of extreme weather is a concept situated at the geographical level. Moreover, emergency decisions related to electric vehicles, mobile energy storage, mobile generators, repair scheduling, and fuel transportation occur within the transportation system at the geographical level. Therefore, incorporating the transportation system as an infrastructure system of equal standing to the energy system and studying the behaviors and performance of the transportation system alongside the energy system under extreme weather conditions is a natural requirement for establishing a more comprehensive assessment framework \cite{Wang_Risk_Unified_2024}. On the one hand, this requires a more detailed consideration of risk compounding effects. For example, although hurricanes may not directly destroy transportation systems, power outages at charging stations caused by distribution line failures could degrade the performance of the transportation system. While rainfall and flooding may have little impact on the power network, they could delay repair work in the transportation network, thereby affecting the repair time for distribution lines and, in turn, impacting power loads. The compound effects of extreme weather on coupled infrastructure systems may significantly impact risk assessments. On the other hand, this also allows for a more thorough exploration of the additional resilience potential in transportation networks, facilitating coordinated decision-making for emergency repairs, emergency power supply, and emergency logistics.

However, the significant heterogeneous difference between energy and transportation systems poses challenges for establishing a comprehensive model \cite{Wang_Risk_Unified_2024}. For example, within the energy system, the same type of energy flow does not exhibit diverse behaviors, only varying in numerical values. However, the state of the transportation system involves a diverse range of vehicle behaviors that are spatially and temporally coupled. Even when vehicles exhibit identical movement characteristics, their behavior types may still vary (e.g., charging/discharging, passenger transport, supply delivery, repairs, or empty movement). The specific behavior type determines the state of the transportation system itself and, due to the interdependence between systems, also affects other systems. In addition, the compound effects of extreme weather may simultaneously impact both energy and transportation systems, introducing new factors into risk assessment. This compounded effect is rarely considered in current resilience research. Even when mobile resources are involved, they often move between power nodes based on static travel time matrices \cite{DING_DRL_Mobile_Restoration_2025}, neglecting dynamic travel delays or road disruptions caused by extreme weather and not treating the transportation system as an entity with its own behavior characteristics.

Furthermore, large-scale urban assessments require models that can accommodate various types of energy-transportation emergency resources to comprehensively reflect risk levels and action value. The energy-transportation coupled system is an interdependent, cross-sector managed entity under attack, and various types of energy-transportation emergency decision resources significantly impact the performance of urban infrastructure during extreme weather events. Accurately integrating diverse emergency decision resources into the model may present scalability challenges. Every additional resource type introduces the need to integrate a new model component into the existing framework, increasing the overall complexity of the model \cite{Wang_Coordination_Restoration_2024, FAN_EI_Restoration_2024}. This leads to the common issue in existing work where, in focusing on a subset of emergency decision combinations as the breakthrough research focus, other subsets may be overlooked. Integrating too many types of energy-transportation emergency decisions into the model makes it difficult to handle. Additionally, models that consider the interdependence of systems will involve multiple entities in energy and transportation, potentially raising privacy concerns \cite{QIU_Hierarchical_MARL_Repair_Crews_2023}. For instance, when the power system assesses its own vulnerabilities, it may not be willing or able to access specific information about the heat network or transportation network.

Building on the above analysis, the main work of this paper is to develop a framework that comprehensively incorporates multiple types of energy-transportation emergency decisions, quantifies risk levels, and identifies vulnerabilities. Compared with the representative studies in Table \ref{table-literature-review-compare}, this paper is dedicated to addressing the challenges posed by the heterogeneous difference between energy and transportation systems, the scalability of accommodating multiple energy-transportation emergency decisions, and cross-sector privacy concerns. 

The contributions of this paper are summarized as follows:

\begin{enumerate}

\item Considering the spatiotemporal evolution of extreme weather, a framework is developed using Monte Carlo simulation with Latin hypercube sampling to quantify the impacts of extreme weather on energy-transportation coupled systems, incorporating multiple sources of uncertainties and balancing probabilities with consequences.

\item The energy and transportation sides, which exhibit heterogeneous differences, are integrated using the spatiotemporally augmented network flow model, thereby enabling the model to fully accommodate multiple types of energy-transportation emergency decisions. Additionally, the model captures the compound impacts of extreme weather on both energy and transportation systems simultaneously at the spatiotemporal level.

\item A method is proposed for targeted identification of system vulnerabilities. Surrogate models based on neural networks and their exact linearized representations are developed to accelerate the process and preserve privacy while maintaining the consideration of interdependence between the coupled infrastructure systems.

\end{enumerate}

The remainder of this paper is organized as follows. Section \ref{Evaluation Framework} establishes the extreme weather risk assessment framework using Monte Carlo simulation. Section \ref{Models} presents each model component needed to construct the integrated energy-transportation model, describing the interactions between infrastructure systems and extreme weather events. Section \ref{Weakness} proposes a method for identifying system vulnerabilities and achieving privacy protection based on neural network surrogates. Section \ref{Numerical Experiments} provides numerical experiments, including case studies based on a real-world city-level system. Section \ref{Conclusion} summarizes the paper.

\section{Risk assessment framework}\label{Evaluation Framework}

Risk assessment under extreme weather differs from traditional power grid security analysis because extreme weather is a spatial concept with time-varying characteristics. The impact of extreme weather's spatiotemporal trajectory on the coupled energy-transportation system differs from traditional fault sets and fault scanning patterns \cite{Wang_Risk_Unified_2024}, as the risks and vulnerabilities are specific to each extreme weather event. Additionally, comprehensive risk assessment cannot assume that failures have already occurred without considering their sources and evolution, which significantly distinguishes it from work focused only on post-disaster recovery optimization \cite{LIBODA2021_RoutingBus}. Multiple types of energy-transportation emergency decisions may occur at different stages of an extreme weather event, necessitating modeling of the spatiotemporal distribution of extreme weather and its physical impact on infrastructure.

\begin{figure}[!t]
	\centering
	\includegraphics[scale=0.11]{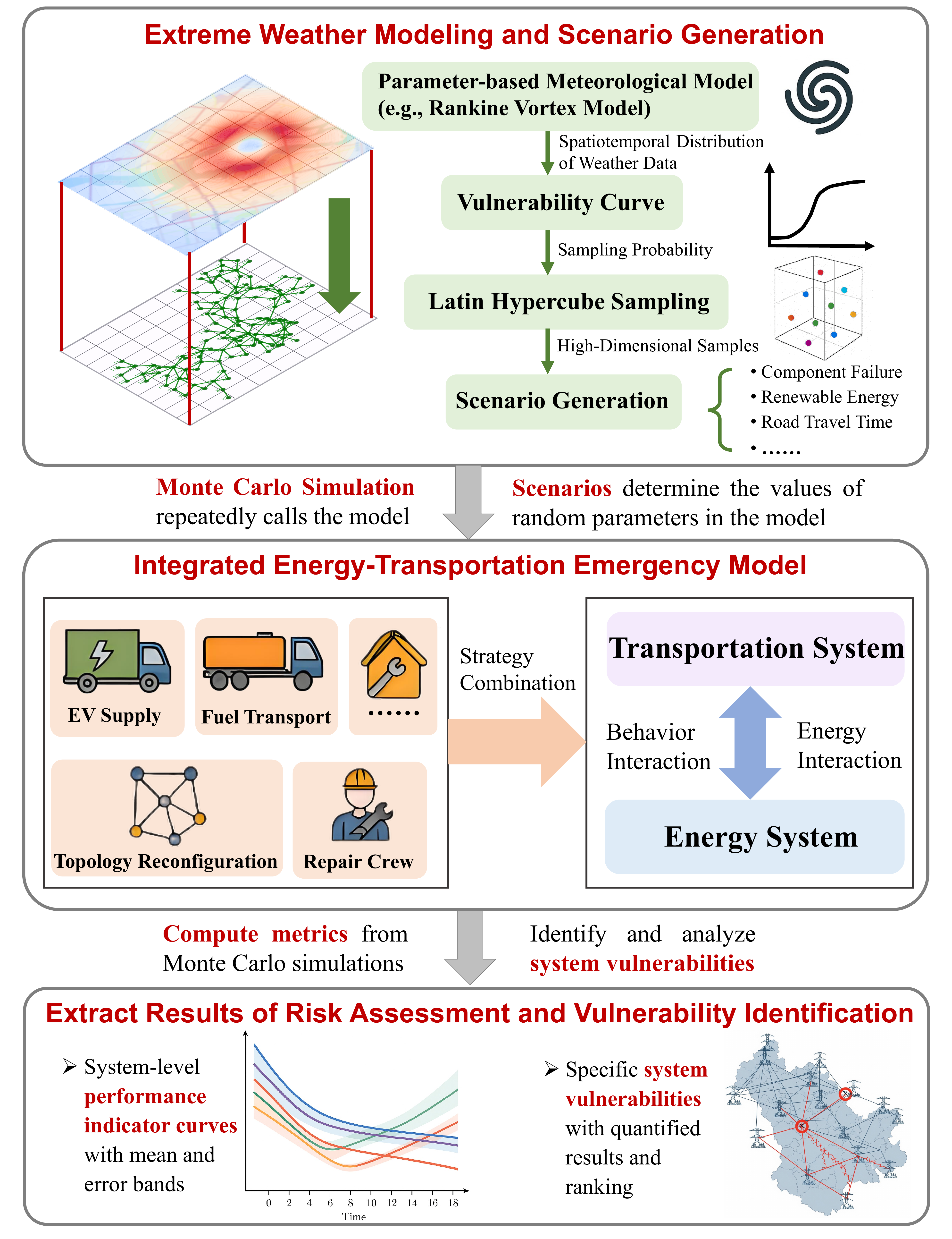}
	\caption{A framework for risk assessment under extreme weather integrating multiple types of energy-transportation emergency decisions.}
	\label{fig-evaluation-framework-new}
\end{figure}

For specific extreme weather events, a risk assessment framework is proposed to simulate and quantify the impact of these events on energy-transportation coupled infrastructure systems, considering the spatiotemporal distribution of extreme weather, accommodating multiple types of energy-transportation emergency decisions, and integrating uncertainties from various sources, as shown in Fig. \ref{fig-evaluation-framework-new}.

Given the known basic parameters and trajectories of meteorological events, the spatiotemporal distribution of weather intensity within a geographic space can be computed using parameter-based meteorological models. For example, in a hurricane event, the wind field within the hurricane's affected area can be simulated using a parameterized wind model such as the Rankine vortex model \cite{PHADKE2003_wind}. The topology of the energy-transportation system must be aligned with the spatiotemporal distribution of extreme weather in a gridded format, allowing the infrastructure to perceive weather intensity over time. Based on this, vulnerability curves serve as a bridge connecting weather intensity to infrastructure performance \cite{Bennett_Nature_extending_2021}. They map the time-varying weather intensity perceived by the infrastructure to compound impacts about performance loss, such as the damage probability of power lines during a hurricane \cite{LIAN2023_cascading_failure_graph,Panteli_Power_System_Resilience_2017} or the effect of rainfall and flooding on road accessibility \cite{PREGNOLATO_impact_road_2017}.

In addition to the damage or performance loss caused by weather intensity, the system still faces multiple uncertainties during extreme weather, including renewable energy output. This paper uses Latin hypercube sampling (LHS) to accommodate random variables from various sources, whether discrete or continuous, and following different probability distributions. LHS is an effective technique for generating high-dimensional uniform samples over $[0,1]$, with its special stratified approach ensuring the generated samples cover the sample space as uniformly and comprehensively as possible. It can be proven that for a random variable $X$ following a uniform distribution on $[0,1]$, the cumulative distribution function of the random variable $F^{(-1)}(X)$ is exactly $F$. Therefore, by performing an inverse function transformation on the samples obtained via LHS, samples from any target distribution can be generated, allowing for the handling of multiple uncertainties.

This paper applies Monte Carlo simulation to quantify system performance and risk using the energy-transportation unified emergency model developed in Section \ref{Models}. The results from the risk assessment framework, whether curves or metrics, are presented in the form of expected values and confidence intervals, thus providing an intuitive display of the severity of specific extreme weather events. The characteristic of the model developed in this paper is its integration of heterogeneous energy and transportation sides in a unified form of network flow, accommodating various combinations of energy-transportation emergency decisions. Any energy-transportation emergency decision can be adapted by the model to either participate or exit. Therefore, Monte Carlo simulations can be repeated under different settings and emergency decision participations, assessing the value of emergency strategies in enhancing resilience, thus providing a benchmark for identifying targeted system vulnerabilities. Furthermore, the model captures the compound impacts of the extreme weather event on both the energy and transportation systems across time and space, enabling a more accurate quantification of the risks associated with such events.

\section{Energy-Transportation System Modeling}\label{Models}

This paper focuses on the power-heat-transportation coupled infrastructure system as a representative research object. It includes both the interdependencies within the energy system (e.g., electricity-heat coupling) and the interdependencies between the energy system and the transportation system. The interdependencies and coupling modeling within the energy system have been well established through energy flow-based methods \cite{Yang_EI_electricity_heat_hydrogen_2024, Xue_Heat_Energy_Flow_2021}. This paper focuses on how to integrate the heterogeneous energy side and the transportation side in a unified network flow format. In extreme weather scenarios, various emergency decisions on the transportation side are key to capturing interdependencies between systems. Tasks like emergency repairs, emergency power supply, fuel transportation, and variable multi-segment travel patterns cannot be fully described by conventional models of electric vehicle users and traffic flows in power-transportation coupling studies \cite{GE2019_CoOptimization_MPCE}.

\begin{figure}[!t]
	\centering
	\includegraphics[scale=0.076]{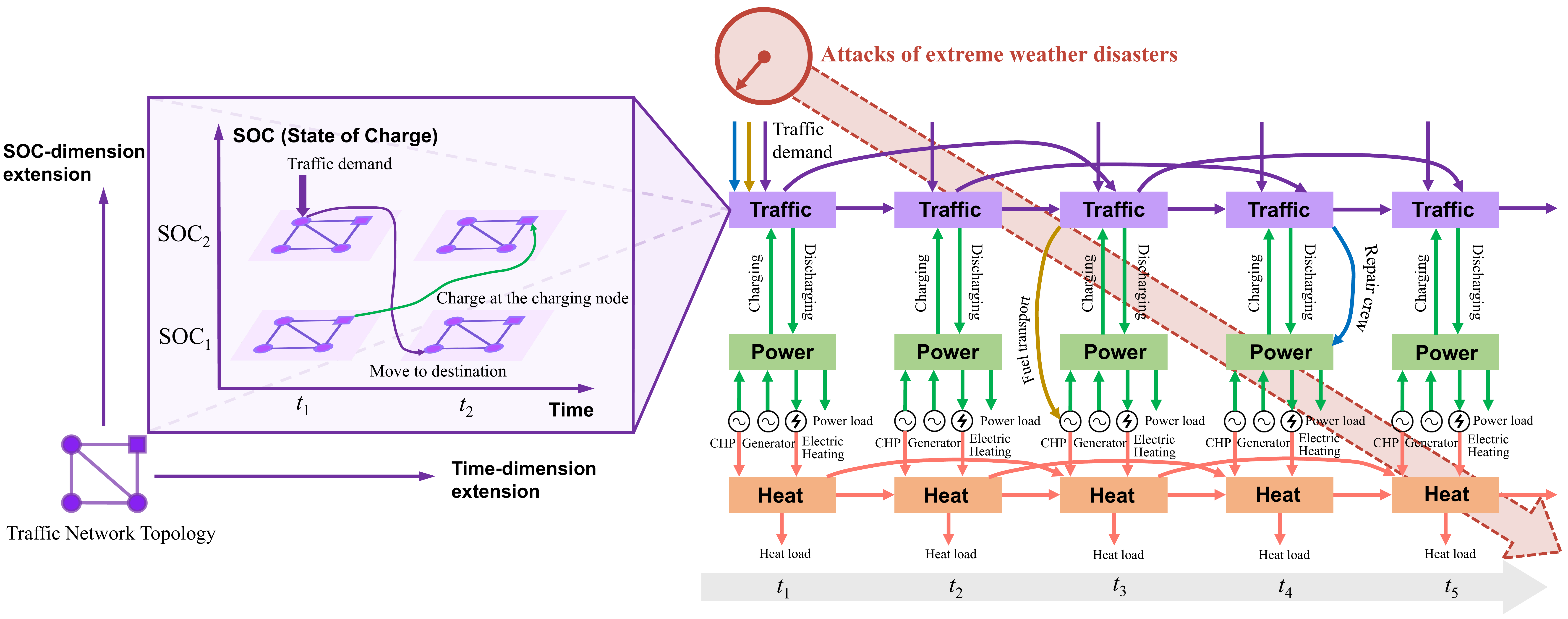}
	\caption{Overall schematic of the power-heat-transportation system model under extreme weather based on network flow.}
	\label{fig-Power-Heat-Traffic-Model}
\end{figure}

This paper applies a spatiotemporal network approach \cite{Wang_AMOD_2024} to extreme weather scenarios and extends it to the energy-transportation system as the modeling foundation. Taking the power-heat-traffic coupled system as an example, as shown in Fig. \ref{fig-Power-Heat-Traffic-Model}, both energy flows and traffic flows move through the nodes and edges of the augmented network. Interdependent networks with different timescales and characteristics can be integrated together. Various complex energy-transportation emergency decisions can be expressed in this integrated model using a unified network flow form, making it possible to fully incorporate combinations of resilience strategies. At the same time, the compound effects of extreme weather on the coupled system in terms of time and space can be captured by the model, such as the simultaneous impact of a hurricane and rainfall on power equipment and transportation roads, reflected in power topology reconfiguration and dynamically varying road travel delays.

The detailed construction of the model in Fig. \ref{fig-Power-Heat-Traffic-Model} is described in the following sections.

\subsection{Energy system}\label{Model-Energy}

The goal of the energy system during extreme weather events is to meet energy supply demands as much as possible (or minimize the loss of energy services) \cite{Wang_Coordination_Restoration_2024, Panteli_Resilience_ProcIEEE_2017}. Additionally, given that critical energy loads should be prioritized \cite{ZHOU2024_Resilience_IEHS}, this paper uses the economic value of load as a performance metric, considering the differences in load importance \cite{LIBODA2022_LTT}.

Objective functions (\ref{Power-obj}) and (\ref{DHN-obj}) represent the economic value of load in the power network and the economic value of load in the heat network, respectively.

\begin{equation}\label{Power-obj}
	F^{\text{PN}}=\sum_{t\in\mathcal{T}}\sum_{i\in\mathcal{N}}\sigma_i^{\text{power}}P^{(l*)}_{i,t}\Delta t
\end{equation}

\begin{equation}\label{DHN-obj}
	F^{\text{HN}}=\sum_{t\in\mathcal{T}}\sum_{i\in\mathcal{H}^{\text{load}}}\sigma_i^{\text{heat}}H^{(l)}_{i,t}\Delta t
\end{equation}

Energy flow in the power network is described by the DC optimal power flow model or the LinDistFlow model \cite{LIBODA2022_PreallocationBus}. The network topology is represented by a series of 0-1 variables, which can be used for topology reconfiguration decisions and are also influenced by extreme weather events. The energy flow in the heat network is modeled using an approximate energy flow model (in \cite{Xue_Heat_Energy_Flow_2021}), with available heat in the pipes (the product of temperature and mass flow rate) as decision variables, and heat losses are considered. 

The above models retain their ability to capture system-level trends. Their accuracy and properties have been analyzed and validated in applications such as post-disaster recovery \cite{LIBODA2022_LTT} and heat-network reconfiguration \cite{Xue_Heat_Energy_Flow_2021}. Due to space limitations, the constraints of the power flow model and the heat network's energy flow model are omitted. Detailed model descriptions and quantitative analyses can be found in the documentation in \cite{Wang_case_data_2024}.

For extreme-weather events, the primary concern is the distribution and supply trends of energy rather than exact operational dynamics. Moreover, factors outside the models, such as prediction accuracy, also affect the results. Thus, although the framework can in principle incorporate precise and complex models, a balance between model complexity and computational performance remains necessary.

\subsection{Transportation system}\label{Model-Traffic}

As shown in the transportation network part of Fig. \ref{fig-Power-Heat-Traffic-Model}, the network flow-based transportation system model is built on an augmented time-space-energy network. The expansion of time, space, and state enables the augmented network to model various complex types of mobility. Spatial movement, time delays, and vehicle state changes are described by the edges between nodes in the augmented network. 

The purpose of introducing transportation modeling is not to generate precise optimization-based emergency and recovery decisions, but to study how to better assess risks and identify vulnerability when such emergency decisions are taken into account. The focus is therefore on capturing overall impacts and trends \cite{Wang_AMOD_2024}, rather than modeling specific microscopic instructions \cite{WANG_mobile_charging}. As a linear-relaxation model that characterizes fleet-dispatch decisions and variable mobility, the network-flow model is sufficiently capable of capturing the essential trends, and its accuracy and practicality have been analyzed and supported in related studies \cite{Estandia2021_AMOD_OPF}.

In the augmented network, $\mathcal{E}$ is the set of nodes. A node is a triplet $\left(n_{\text{loc}},n_{\text{time}},n_{\text{soc}}\right)$ representing space, time, and state of charge (SOC) attributes. $i_{\text{ori},\sim}$ and $i_{\text{des},\sim}$ represent the start and end nodes' attributes of edge $i$ in the augmented network. Additionally, $i_{\text{dist}}$ represents the spatial distance of edge $i$ in the augmented network. $i_\text{delay}$ is the additional time delay number corresponding to edge $i$, indicating that traveling along edge $i$ to reach the destination takes an additional $i_\text{delay}$ units of $\Delta t$ compared to the ideal scenario.

The goal of the transportation system during extreme weather events is to meet mobility demands while minimizing time delays. Objective function (\ref{Traffic-obj}) represents the economic value of traffic demand in the transportation network, subtracting the time delay cost.

\begin{equation}\label{Traffic-obj}
	\begin{aligned}
	F^{\text{TN}}=\sum_{i\in \mathcal{I}^{\text{SER}}}x^{\text{SER}}_i\left(r^{\text{in},(\text{0})}_i+r^{\text{in},\text{(1)}}_i i_\text{dist}-r_i^{\text{time}}i_\text{delay}\Delta t\right)
	\end{aligned}
\end{equation}

To comprehensively describe the multiple types of behaviors in the transportation system, different types of edges and traffic flows are defined. $\mathcal{I}^{\text{type}}, x_i^{\text{type}}$ represent the set of edges of type ``type'' and the traffic flow on edge $i$ of that type. In this context, $\text{type} \in {\text{SER, RE, CHA, DIS, STOP, GR, IR, FT}}$ includes the following: SER (satisfying traffic demand), RE (empty vehicle movement), CHA (charging), DIS (discharging), STOP (remaining stationary), GR (emergency repair movement), IR (in the process of repairing), and FT (fuel transport).

\begin{subequations}
	\begin{align}
	&\forall x_i^{\text{type}},x_{\text{ori},\text{des},t}^{\text{drop}},P^{\text{cha}}_{f,t},P^{\text{dis}}_{f,t},\xi^{\text{sup}}_{s,t},\theta_{g,t},e_n\geq \text{0}
	\label{Traffic-0}\\
	&\begin{aligned}
	\sum_{\text{type}}\left(\sum\limits_{\substack{i\in \mathcal{I}^{\text{type}}\\i_{\text{ori}}=n}}x^{\text{type}}_i-\sum\limits_{\substack{i\in \mathcal{I}^{\text{type}}\\i_{\text{des}}=n}}x^{\text{type}}_i\right)
	=s_n-e_n,\ \forall n\in \mathcal{E}
	\end{aligned}
	\label{Traffic-2}\\
	&e_n=\text{0},\ \forall n\in \mathcal{E},\text{if}\ n_{\text{time}}<t^{\text{max}}\  \text{or}\ n_{\text{soc}}<c^{\text{0}}
	\label{Traffic-8}\\
	&\sum_{n\in \mathcal{E}}e_n=\sum_{n\in \mathcal{E}}s_n
	\label{Traffic-9}\\
	&\begin{aligned}
	\sum\limits_{\substack{i\in \mathcal{I}^{\text{SER}}\\
	i_{\text{ori,loc}}=\text{ori}\\
	i_{\text{des,loc}}=\text{des}\\
	i_{{\text{ori,time}}}=t
	}}
	x^{\text{SER}}_i=o_{\text{ori},\text{des},t}-x^{\text{drop}}_{\text{ori},\text{des},t},\ 
	\forall \text{ori},\text{des}\in \mathcal{G},\forall t\in \mathcal{T}
	\end{aligned}
	\label{Traffic-1}\\
	&C^{\text{cha,max}}_f\geq P_{f, t}^{\text{cha}}=P^{\text{cha}}\sum\limits_{\substack{i\in \mathcal{I}^{\text{CHA}}\\i_{\text{ori,loc}}=f\\i_{\text{ori,time}}=t}}x^{\text{CHA}}_i,\ \forall f\in \mathcal{F}, \forall t\in \mathcal{T}
	\label{Traffic-3}\\
	&C^{\text{dis,max}}_f\geq P_{f, t}^{\text{dis}}=P^{\text{dis}}\sum\limits_{\substack{i\in \mathcal{I}^{\text{DIS}}\\i_{\text{ori,loc}}=f\\i_{\text{ori,time}}=t}}x^{\text{DIS}}_i,\ \forall f\in \mathcal{F}, \forall t\in \mathcal{T}
	\label{Traffic-4}\\
	&\xi_{s, t}^{\text{sup}}=\sum\limits_{\substack{i\in \mathcal{I}^{\text{FT}}\\i_{\text{des,loc}}=s\\i_{\text{des,time}}=t}}x^{\text{FT}}_i,\ \forall s\in \mathcal{S'}, \forall t\in \mathcal{T}
	\label{Traffic-5}\\
	&\theta_{g, t}=\sum\limits_{\substack{i\in \mathcal{I}^{\text{IR}}\\i_{\text{ori,loc}}=g\\i_{\text{ori,time}}=t}}x^{\text{IR}}_i,\ \forall g\in \mathcal{G}, \forall t\in \mathcal{T}
	\label{Traffic-6}\\
	&\sum\limits_{\substack{i\in \mathcal{I}^{\text{IR}}\\i_{\text{ori}}=n}} x_i^{\text{IR}}\leq \sum\limits_{\substack{i\in \mathcal{I}^{\text{GR}}\\i_{\text{des}}=n}}x_i^{\text{GR}} + \sum\limits_{\substack{i\in \mathcal{I}^{\text{IR}}\\i_{\text{des}}=n}}x_i^{\text{IR}},\ \forall n\in \mathcal{E}
	\label{Traffic-7}
	\end{align}
	\label{Traffic}
\end{subequations}

Constraints (\ref{Traffic-0})-(\ref{Traffic-9}) define the basic behavior of the network flow. Constraint (\ref{Traffic-0}) is the non-negativity constraint. Constraint (\ref{Traffic-2}) is the network flow conservation at the nodes. Constraints (\ref{Traffic-8}) and (\ref{Traffic-9}) combined require that all flows must exit the network from a node with SOC of at least $c^{\text{0}}$ at the final time period $t^{\text{max}}$, while also ensuring that the total number of vehicles remains constant. This means that, at the end of the time period, all vehicles should have an SOC no less than $c^{\text{0}}$.

On this basis, additional descriptions are added to reflect the impact of each type of vehicle activity. Constraint (\ref{Traffic-1}) describes the behavior that satisfies traffic demand. Constraints (\ref{Traffic-3}) and (\ref{Traffic-4}) aggregate charging and discharging vehicles into charging and discharging power. Constraint (\ref{Traffic-5}) aggregates fuel supply vehicles into the fuel supply amount. Constraint (\ref{Traffic-6}) aggregates repair vehicles into the number of resources committed to the repair work. Constraint (\ref{Traffic-7}) describes the continuity of the repair process.

Through the above constraints, vehicle mobility in the transportation system is represented as different types of network flows, which move separately through the augmented network. Charging and discharging, emergency repairs, and fuel scheduling all manifest as unified mobility in the transportation system. However, these task-driven traffic flows will induce cross-system coupling effects at their spatial-temporal destinations upon reaching their end nodes.

\begin{figure}[!t]
	\centering
	\includegraphics[scale=0.145]{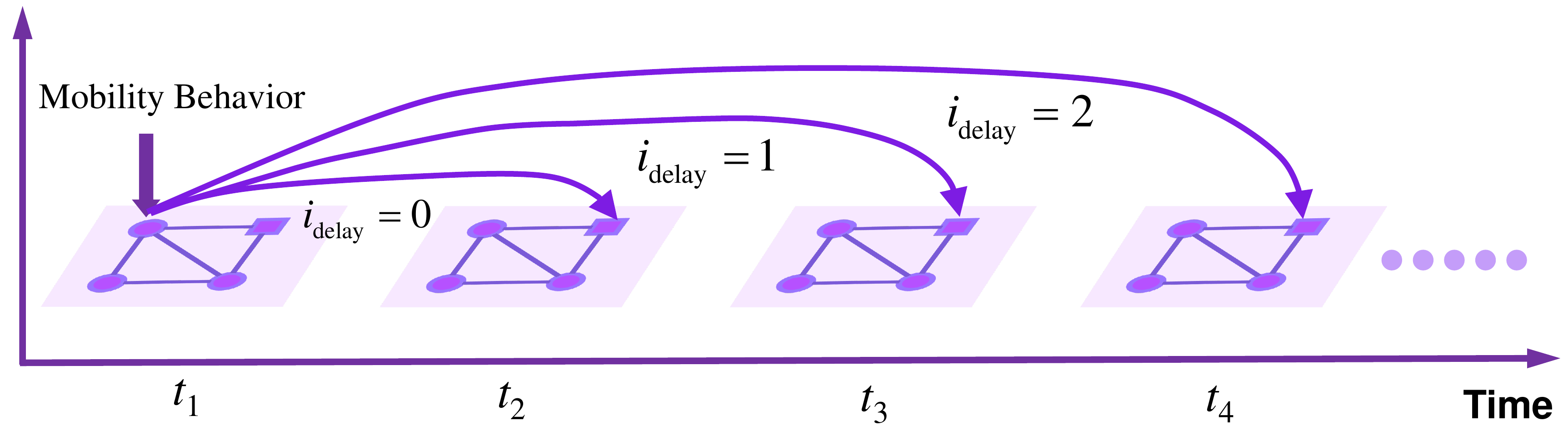}
	\caption{Schematic of edges with different time delays for the same mobility behavior in the spatiotemporal augmented network.}
	\label{Mobility-Delay}
\end{figure}

Traditional spatio-temporal augmented networks are mostly based on static travel time matrices between nodes \cite{DING_DRL_Mobile_Restoration_2025, YAN_VPP_Mobile_Storage_2025}, which fail to account for the dynamic effects of extreme weather, including road interruption. In this paper, this effect is accommodated by adding redundant edges in the augmented network, as shown in Fig. \ref{Mobility-Delay}. For the same spatial movement, using the required travel time under ideal conditions as a baseline, edges with different time spans, differentiated by $i_\text{delay}$, are introduced to represent potential delays in travel time on the roads. Thus, different travel times required for the same movement behavior under different extreme weather conditions can be represented by selecting these different edges. Constraints (\ref{Attack-Traffic}) enforce this selection process of the network flow.

\begin{subequations}
	\begin{align}
	&\begin{aligned}
	x_i^{\text{type}}=0, &\text{if}\ \prod_{(a,b)\in\text{path}(i)}u_{a,b,t_{i,a}}=0,\ \\
	&\forall i \in \mathcal{I}^{\text{type}}, \text{type}\in\{\text{SER,RE,GR,FT}\}
	\end{aligned}
	\label{Attack-Traffic-1}\\
	&\begin{aligned}
	x_i^{\text{type}} = 0, &\text{if}\ i_{\text{delay}}\neq\left\lfloor\frac{\text{1}}{\Delta t}\sum_{(a,b)\in\text{path}(i)}\frac{\text{1}-v_{a,b,t_{i,a}}}{v_{a,b,t_{i,a}}}T_{a,b}\right\rceil ,\ \\
	&\forall i \in \mathcal{I}^{\text{type}}, \text{type}\in\{\text{SER,RE,GR,FT}\}
	\end{aligned}
	\label{Attack-Traffic-2}
	\end{align}
	\label{Attack-Traffic}
\end{subequations}

Constraint (\ref{Attack-Traffic-1}) ensures that all movements (traffic demand movement, empty vehicle movement, emergency repairs, fuel transport) require that the roads in the selected path are not interrupted. Constraint (\ref{Attack-Traffic-2}) chooses out the time delays of the path corresponding to each movement behavior, requiring the $i_\text{delay}$ of the selected edge $i$ to satisfy the current road conditions, while limiting the flow on other redundant edges corresponding to the same movement behavior to zero.

\subsection{System coupling}\label{Coupling}

The coupling between the energy system and the transportation system can be summarized in energy, logistics, and repair aspects.

First, in terms of energy, for a coupled node $i$ in the power network, its load includes both electric heating power and charging/discharging power, as shown in equation (\ref{coupling-cons-power-load}).

\begin{equation}
	P^{(l)}_{i,t}=P^{(l*)}_{i,t}+\frac{H^{(g)}_{i,t}}{\mu_i}+P_{i, t}^{\text{cha}}-P_{i, t}^{\text{dis}},\ \forall i\in\mathcal{N},\forall t\in\mathcal{T}
	\label{coupling-cons-power-load}
\end{equation}

Additionally, coupling constraints for devices within the energy system may also be introduced based on specific conditions, such as characterizing the feasible operational region of combined heating and power (CHP) units using a convex combination of the extreme points \cite{Xue_Heat_Energy_Flow_2021}.

In terms of logistics, for fuel-consuming units, the fuel reserves and consumption are jointly affected by their electric and heat output, and they may also receive additional supply from transportation behaviors, as shown in equation (\ref{coupling-cons-fuel}).

\begin{equation}
	\begin{aligned}
	0\leq\xi_{s,t+1}=&\xi_{s,t}-\Delta t\frac{P^{(g)}_{s,t}+H^{(g)}_{s,t}}{\eta^{\text{uti}}_s} + \xi_{s,t}^{\text{sup}},\ \\
	&\forall s\in\mathcal{S'},\forall t\in\mathcal{T}\setminus\{t^\text{max}\}
	\end{aligned}
	\label{coupling-cons-fuel}
\end{equation}

Finally, regarding repair actions, the following coupling constraints are applied. 

\begin{subequations}
	\begin{equation}
	\theta_{j,t}\geq R^{\text{need}}_{i,j}\delta_{i,j,t},\ \forall (i,j)\in\mathcal{B},\forall t\in\mathcal{T}
		\label{coupling-cons-repair-1}
	\end{equation}
	\begin{equation}
	\begin{aligned}
	s_{i,j,t-1}\leq s_{i,j,t}\leq &s_{i,j,t-1} + \prod_{\tau=t-t^{\text{repair}}_{i,j}}^{t-1}\delta_{i,j,\tau},\\
	&\forall (i,j)\in\mathcal{B},\forall t\in \mathcal{T},t-t^{\text{repair}}_{i,j}\geq t_{i,j}^{\text{break}}
	\end{aligned}
	\label{coupling-cons-repair-2}
	\end{equation}
	\begin{equation}
	\begin{aligned}
	s_{i,j,t}=0,\ \forall (i,j)\in\mathcal{B},\forall t\in \mathcal{T},t_{i,j}^{\text{break}}\leq t<t_{i,j}^{\text{break}}+t^{\text{repair}}_{i,j}
	\end{aligned}
	\label{coupling-cons-repair-3}
	\end{equation}
	\label{Repair}
\end{subequations}

Constraint (\ref{coupling-cons-repair-1}) ensures that if $\delta_{i,j,t} = 1$, the repair resources for line $(i,j)$ must exceed the threshold. Constraints (\ref{coupling-cons-repair-2}) and (\ref{coupling-cons-repair-3}) require that if a destroyed line can be re-closed, there must have been a continuous repair process with sufficient resources for $t^{\text{repair}}_{i,j}$ periods before the closure, or it must have already been repaired and closed in earlier periods. The required repair time $t^{\text{repair}}_{i,j}$ may be a random variable \cite{Zhang_IET_Assessment_2019}.

\subsection{Model summary}\label{Model-Summary}

After predicting extreme weather events, the principle of maximizing social welfare is considered. The centralized optimization of the energy-transportation coupling system is used as a tool to evaluate the risks faced by the infrastructures. Through the modeling procedure above, the physical metrics of each subsystem are converted into economic values or equivalent social utilities, enabling an integrated assessment of economic losses caused by extreme weather across coupled systems. This treatment aligns with the common practice used in cross-system objective formulation \cite{CUI_EV_routing_wireless_power_restoration_2026}.

The compact form of the complete model is shown in equation (\ref{Model-Complete-Form}).

\begin{equation}
	\begin{aligned}
	\max\ &\text{Obj}=F^{\text{PN}}\left(\mathbf{x}^{\text{PN}}\right)+F^{\text{HN}}\left(\mathbf{x}^{\text{HN}}\right)+F^{\text{TN}}\left(\mathbf{x}^{\text{TN}}\right)\\
	\text{s.t.}\ &\mathbf{g}^{\text{PN}}\left(\mathbf{x}^{\text{PN}}\right)\leq \mathbf{0},\ \mathbf{h}^{\text{PN}}\left(\mathbf{x}^{\text{PN}}\right)=\mathbf{0}\\
	&\mathbf{g}^{\text{HN}}\left(\mathbf{x}^{\text{HN}}\right)\leq \mathbf{0},\ \mathbf{h}^{\text{HN}}\left(\mathbf{x}^{\text{HN}}\right)=\mathbf{0}\\
	&\mathbf{g}^{\text{TN}}\left(\mathbf{x}^{\text{TN}}\right)\leq \mathbf{0},\ \mathbf{h}^{\text{TN}}\left(\mathbf{x}^{\text{TN}}\right)=\mathbf{0}\\
	&\mathbf{h}^{\text{CO}}\left(\mathbf{x}^{\text{PN}}_2,\mathbf{x}^{\text{HN}}_2,\mathbf{x}^{\text{TN}}_2\right)=\mathbf{0}\\
	&\mathbf{g}^{\text{CO}}\left(\mathbf{x}^{\text{PN}}_2,\mathbf{x}^{\text{TN}}_2\right)\leq\mathbf{0}
	\end{aligned}
	\label{Model-Complete-Form}
\end{equation}

Let $\mathbf{x}^{\text{PN}}=\left[\mathbf{x}^{\text{PN}}_{1}, \mathbf{x}^{\text{PN}}_2\right]$ represents the decision variables related to the power system, where $\mathbf{x}^{\text{PN}}_2=\left[P_{i,t}^{(l)};P_{i,t}^{(l*)};\delta_{i,j,t};\xi_{s,t}\right]$ represents the variables coupled between the power system and other systems. Let $\mathbf{x}^{\text{HN}}=\left[\mathbf{x}^{\text{HN}}_{1}, \mathbf{x}^{\text{HN}}_2\right]$ represents the decision variables related to the heat system, where $\mathbf{x}^{\text{HN}}_2=\left[H_{i,t}^{(g)}\right]$ represents the variables coupled between the heat system and other systems. Let $\mathbf{x}^{\text{TN}}=\left[\mathbf{x}^{\text{TN}}_{1}, \mathbf{x}^{\text{TN}}_2\right]$ represents the decision variables related to the transportation system, where $\mathbf{x}^{\text{TN}}_2=\left[P_{i,t}^{\text{cha}};P_{i,t}^{\text{dis}};\xi_{s,t}^{\text{sup}};\theta_{j,t}\right]$ represents the variables coupled between the transportation system and other systems.

In model (\ref{Model-Complete-Form}): $\mathbf{g}^{\text{PN}}$ and $\mathbf{h}^{\text{PN}}$ represent the power flow model and the topology constraints. $\mathbf{g}^{\text{HN}}$ and $\mathbf{h}^{\text{HN}}$ represent the energy flow model for the heat network and the related constraints. $\mathbf{g}^{\text{TN}}$ and $\mathbf{h}^{\text{TN}}$ represent the transportation system constraints (\ref{Traffic}) and (\ref{Attack-Traffic}). The coupling constraint $\mathbf{h}^{\text{CO}}$ involves constraints (\ref{coupling-cons-power-load}) and (\ref{coupling-cons-fuel}). The coupling constraint $\mathbf{g}^{\text{CO}}$ involves constraint (\ref{Repair}).

The computational complexity of the full model depends on the temporal resolution and the scale of nodes and edges in each system. For the power network, the number of variables and constraints grows on the order of $\mathcal{O}\left(\left|\mathcal{N}\right|\cdot \left|\mathcal{T}\right|\right)+\mathcal{O}\left(\left|\mathcal{B}\right|\cdot \left|\mathcal{T}\right|\right)$. For the heat network, it grows on the order of $\mathcal{O}\left(\left|\mathcal{H}\right|\cdot \left|\mathcal{T}\right|\right)+\mathcal{O}\left(\left|\mathcal{P}\right|\cdot \left|\mathcal{T}\right|\right)$. For the transportation network, it grows on the order of $\mathcal{O}\left(\left|\mathcal{E}\right|\right)+\mathcal{O}\left(\displaystyle\sum_{\text{type}}\left|\mathcal{I}^{\text{type}}\right|\right)$.

\section{Identification of vulnerable components and surrogate-based methods}\label{Weakness}

The proposed risk assessment framework enables overall-level evaluation of extreme weather impacts, offering strategic insights for cross-sector decision-making bodies. Furthermore, identifying and quantifying system vulnerabilities under such risks allows for a deeper understanding of how component failures contribute to objective losses, thereby providing practical guidance for targeted prevention and hardening strategies \cite{Sturmer_Hardening_Lines_2024}.

\subsection{Direct indicator for quantifying vulnerable components}\label{Direct Assessment}

Vulnerability reflects the distribution and ranking of weak components within the system, highlighting those most worth prioritizing for reinforcement. The degree of vulnerability integrates both the component's failure probability and the full consequences of its failure. These consequences depend not only on the component's criticality within its own system but also on the potential cross-system losses it may cause, all of which are influenced by complex interactions with multi-type energy-transportation emergency decisions. Due to this complexity, existing vulnerability indicators based on graph theory or physical characteristics of power networks are difficult to transfer easily and reasonably to energy-transportation coupled systems, and heuristic indicators inherently involve some subjectivity.

To fully reflect the concept of vulnerability in this context, this paper advocates for a simulation-oriented, direct indicator. As defined in equation (\ref{weakness_definition}), this indicator evaluates the vulnerability of component $b$ by comparing the system's expected objective value in two cases: before reinforcement ($\mathbb{E}\left(\text{Obj}_0\right)$) and after reinforcement ($\mathbb{E}\left(\text{Obj}_b\right)$). In the ``before'' case, the component follows the original fragility probability distribution, and in the ``after'' case, the component is assumed to be hardened and immune to damage (or its fragility curve can be scaled according to practical needs \cite{Panteli_Resilience_ProcIEEE_2017}). The difference between the two cases represents the utility $U_b$ brought by reinforcing component $b$.

\begin{equation}\label{weakness_definition}
	U_{b} = \mathbb{E}\left(\text{Obj}_b\right)-\mathbb{E}\left(\text{Obj}_0\right)
\end{equation}

This indicator inherently accounts for complex elements such as probabilistic uncertainty, interdependencies across systems, and multiple types of energy-transportation emergency decisions, which are embedded within the unified energy-transportation emergency model and risk assessment framework. It is clearly defined and free from heuristic subjectivity. Moreover, since the indicator directly simulates the expected outcome of reinforcement actions, it provides the most direct basis for targeted preventive actions.

\subsection{Neural network surrogate-based method}

The direct computation of equation (\ref{weakness_definition}) faces a major challenge due to the computational burden of repeatedly running Monte Carlo simulations. To rank the components by weakness and identify the most critical ones, full-scale simulations must be conducted for both the ``before reinforcement'' and ``after reinforcement'' states of each component, which is computationally intensive. Moreover, identifying vulnerabilities is often the primary concern of individual system operators. Taking power lines as an example, they are assets of the power utility, which is also responsible for their hardening and repair during extreme weather events. Although the centralized model in equation (\ref{Model-Complete-Form}) and the indicator in equation (\ref{weakness_definition}) provide a theoretical foundation, practical implementation may face collaboration barriers. A power utility seeking to assess the vulnerability of its components from the social welfare perspective would need access to models and detailed data from the heat and transportation systems, which may raise privacy concerns. Whether the utility builds and maintains models of other systems or engages in repeated multi-party iterations, both options increase computational or communication burden and demand a higher level of inter-system coordination.

\begin{figure}[!t]
	\centering
	\includegraphics[scale=0.33]{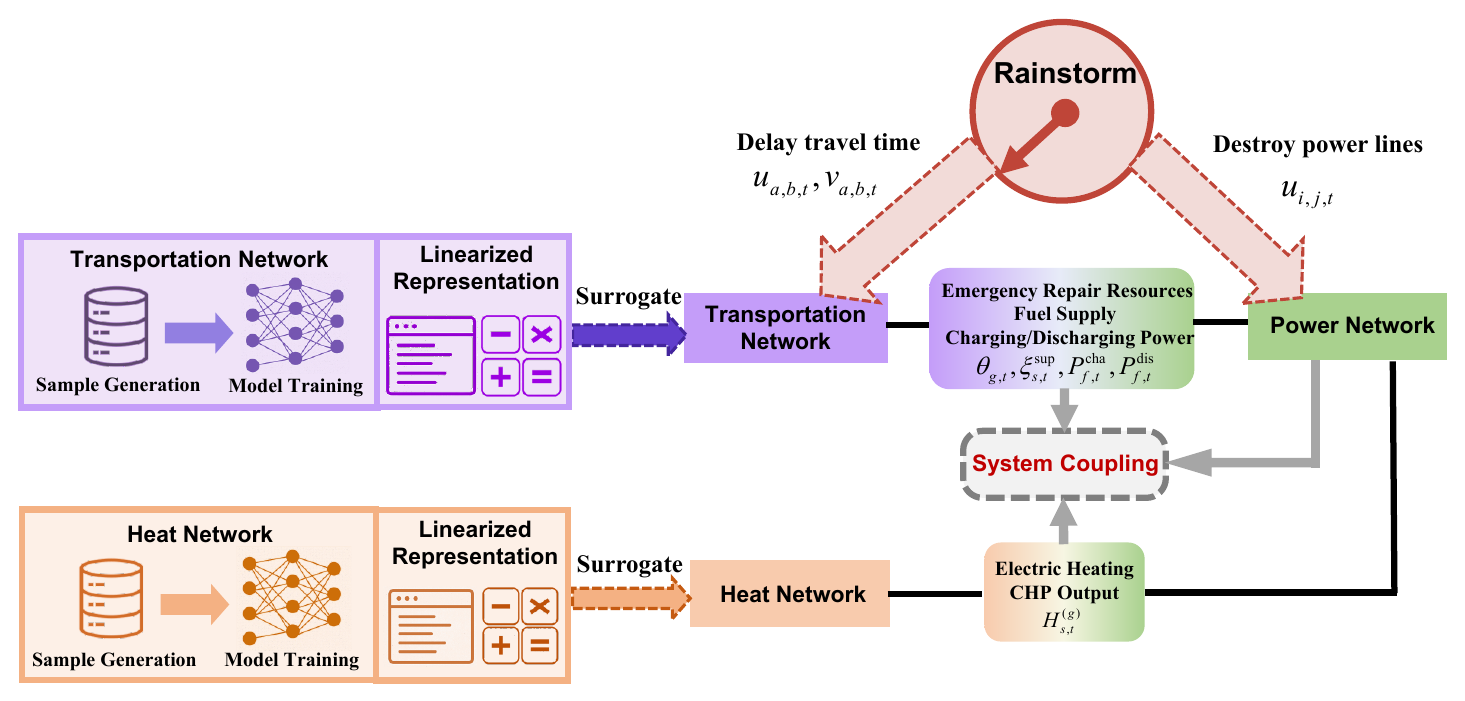}
	\caption{Surrogate-based approach for system vulnerability identification and quantification.}
	\label{surrogate_framework}
\end{figure}

To accelerate and decouple the identification of vulnerable components, this paper proposes a neural network-based surrogate method, as illustrated in Fig. \ref{surrogate_framework}. The core idea is to train neural networks that approximate the mapping from boundary variables to subsystem objective values, allowing the power system operator to use these surrogates to represent interdependencies with the heat and transportation systems. The proposed method involves two stages: training and embedding.

In the training stage, the heat and transportation systems independently generate samples and train their own surrogate models. The input to each model is its respective set of boundary variables ($\widetilde{\mathbf{x}}_{2}^{\text{HN}}$ and $\widetilde{\mathbf{x}}_{2}^{\text{TN}}$), while the output is the system's objective value. The input ranges can be selected based on historical data or physical limits. For example, the heat source output $H_{i,t}^{(g)}$ can be sampled between its minimum and maximum output while ensuring generalization.

By examining the structure of model (\ref{Model-Complete-Form}), it is clear that when $\widetilde{\mathbf{x}}_{2}^{\text{HN}}$ and $\widetilde{\mathbf{x}}_{2}^{\text{TN}}$ are fixed, the heat and transportation systems' objective values depend only on their internal variables. Therefore, they can be decoupled and solved independently, as shown in equations (\ref{Model-Complete-Form-Heat-Sub}) and (\ref{Model-Complete-Form-Traffic-Sub}). These subproblems are much smaller than the full integrated model and do not involve binary variables about power network topology, significantly reducing the computational burden. Furthermore, since this stage is conducted offline, it can be completed in advance by each system.

\begin{subequations}
	\begin{equation}
		\begin{aligned}
		\max\ &F^{\text{HN}}\left(\mathbf{x}_{1}^{\text{HN}}, \widetilde{\mathbf{x}}_{2}^{\text{HN}}\right)\\
		\text{s.t.}\ &\mathbf{g}^{\text{HN}}\left(\mathbf{x}_{1}^{\text{HN}}, \widetilde{\mathbf{x}}_{2}^{\text{HN}}\right)\leq \mathbf{0},\ \mathbf{h}^{\text{HN}}\left(\mathbf{x}_{1}^{\text{HN}}, \widetilde{\mathbf{x}}_{2}^{\text{HN}}\right)=\mathbf{0}\\
		\end{aligned}
		\label{Model-Complete-Form-Heat-Sub}
	\end{equation}
	\begin{equation}
		\begin{aligned}
		\max\ &F^{\text{TN}}\left(\mathbf{x}_{1}^{\text{TN}},\widetilde{\mathbf{x}}_{2}^{\text{TN}}\right)\\
		\text{s.t.}\ &\mathbf{g}^{\text{TN}}\left(\mathbf{x}_{1}^{\text{TN}}, \widetilde{\mathbf{x}}_{2}^{\text{TN}}\right)\leq \mathbf{0},\ \mathbf{h}^{\text{TN}}\left(\mathbf{x}_{1}^{\text{TN}}, \widetilde{\mathbf{x}}_{2}^{\text{TN}}\right)=\mathbf{0}\\
		\end{aligned}
	\label{Model-Complete-Form-Traffic-Sub}
\end{equation}
\label{Model-Complete-Sub-System}
\end{subequations}

By leveraging the fitting capability of neural networks, the surrogate model aims to capture the value and trends of the subsystems, as expressed in equation (\ref{NN-Sub-System}).

\begin{subequations}
	\begin{equation}
		\begin{aligned}
		{F}^{\text{HN}}\left(\mathbf{x}^{\text{HN}}\right)\approx\text{Surrogate}^{\text{HN}}\left(\mathbf{x}_{2}^{\text{HN}}\right)
		\end{aligned}
		\label{NN-Heat-Sub}
	\end{equation}
	\begin{equation}
		\begin{aligned}
		{F}^{\text{TN}}\left(\mathbf{x}^{\text{TN}}\right)\approx\text{Surrogate}^{\text{TN}}\left(\mathbf{x}_{2}^{\text{TN}}\right)
		\end{aligned}
	\label{NN-Traffic-Sub}
\end{equation}
\label{NN-Sub-System}
\end{subequations}

In the embedding stage, the power system, acting as the primary analysis entity, receives the pre-trained surrogate models and incorporates them into its own vulnerability identification process. As shown in equation (\ref{Model-Complete-Form-Power-Sub}), the power system embeds the surrogate model into the complete model (\ref{Model-Complete-Form}), with all internal variables and constraints of the heat and transportation systems eliminated. While the power system remains unaware of the detailed models and parameters of the other subsystems, it can still approximately perceive the interdependencies.

\begin{equation}
	\begin{aligned}
	\max\ &F^{\text{PN}}\left(\mathbf{x}^{\text{PN}}\right)+\text{Surrogate}^{\text{HN}}\left(\mathbf{x}_{2}^{\text{HN}}\right)+\text{Surrogate}^{\text{TN}}\left(\mathbf{x}_{2}^{\text{TN}}\right)\\
	\text{s.t.}\ &\mathbf{g}^{\text{PN}}\left(\mathbf{x}^{\text{PN}}\right)\leq \mathbf{0},\ \mathbf{h}^{\text{PN}}\left(\mathbf{x}^{\text{PN}}\right)=\mathbf{0}\\
	&\mathbf{h}^{\text{CO}}\left(\mathbf{x}^{\text{PN}}_2,\mathbf{x}^{\text{HN}}_2,\mathbf{x}^{\text{TN}}_2\right)=\mathbf{0}\\
	&\mathbf{g}^{\text{CO}}\left(\mathbf{x}^{\text{PN}}_2,\mathbf{x}^{\text{TN}}_2\right)\leq\mathbf{0}
	\end{aligned}
	\label{Model-Complete-Form-Power-Sub}
\end{equation}

Compared to the full centralized model (\ref{Model-Complete-Form}), the surrogate-based model (\ref{Model-Complete-Form-Power-Sub}) significantly reduces the number of variables and constraints, enabling the power system to independently evaluate its own vulnerabilities using this formulation as a tool. The primary challenge in solving model (\ref{Model-Complete-Form-Power-Sub}) lies in the highly non-convex and nonlinear functions introduced by the surrogate model. One solution is to apply nonlinear or black-box optimization algorithms, such as COBYLA (Constrained Optimization BY Linear Approximations) \cite{Powell_trust_optimization_2003}. Alternatively, for specific surrogate model structures, exact mixed-integer reformulations can be achieved using auxiliary variables and constraints \cite{Maragno_constraint_learning_2023}.

A neural network with ReLU activation functions and linear hidden layers is used as a representative. For a given neuron, let $x_1,\cdots,x_n$ be the outputs of the $n$ neurons in the previous layer, $\beta_1,\cdots,\beta_n$ be the weights of the connections from these neurons, and $\beta_0$ be the bias term. The forward propagation process is a linear combination, with the activation function being the only nonlinear component. The ReLU activation function is essentially a $\max$ operator, which can be linearized using the Big-M method, where $M$ is a sufficiently large positive constant and $u$ is a binary auxiliary variable. 

\begin{equation}
	\begin{aligned}
	y=\text{ReLU}(w)=\max\left(0,w\right)\\
	=\max\left(0,\sum_{i=1}^{n}\beta_ix_i+\beta_0\right)
	\end{aligned}
	\Longrightarrow
	\begin{cases}
		y\geq w,\ y\leq w+uM,\\
		y\geq 0,\ y\leq(1-u)M\\
		u\in\left\{0,1\right\}
	\end{cases}
	\label{agent_embedding_bigM}
\end{equation}

As shown in equation (\ref{agent_embedding_bigM}), the forward propagation process of the neural network surrogate can thus be exactly represented as a set of linearized constraints. This linearization works properly under the condition $M >|w|$, and an upper bound on $|w|$ can be derived from the value ranges of $x_1\sim x_n$. Starting from the known bounds of the physical input variables at the first layer, the bounds of each layer's neuron outputs are derived recursively, enabling the use of tighter Big-M values. This procedure can also be automated through the general constraint features of advanced optimization solvers.

If the neural network contains a total of $\left|\mathcal{X}\right|$ neurons across all layers, embedding it into the model introduces additional variables and constraints on the order of $\mathcal{O}\left(\left|\mathcal{X}\right|\right)$. These constraints focus on capturing the influence trends of other coupled systems, thereby replacing the internal variables and constraints of those systems themselves.

\section{Numerical experiments}\label{Numerical Experiments}

This paper sets up two energy-transportation coupling test systems with different scales and characteristics. 

System\#1 (33-bus power distribution network, 27-node heat network, 33-node transportation network) includes standard systems widely used in academic literature and serves as the main focus of the analysis. The power network includes two CHP units and one photovoltaic (PV) generator. In the heat network, an electric boiler serves as a heat source aside from the CHP units. A typhoon of approximately level 15, accompanied by a rainstorm, is set to make landfall at the city's edge of system\#1 at 12:00 on Day 1 and move west to east. Its spatiotemporal profile is simulated using the modified Rankine vortex model \cite{PHADKE2003_wind}. The vulnerability curves of power lines follow \cite{LIAN2023_cascading_failure_graph, Zhang_IET_Assessment_2019}. The economic value coefficients of loads take reference from \cite{LIBODA2022_LTT, ZHOU2024_Resilience_IEHS}. The nominal average vehicle speed is set to 30 km/h, and the impact of rainfall-induced flooding depth on traffic speed follows the empirical relation in \cite{PREGNOLATO_impact_road_2017}. The simulation covers 48 hours of forecast horizon.

System\#2 is based on real-world data from a city's power network and heat network in China, providing a supplementary example for large-scale systems. Its basic settings are similar to system\#1, but several adjustments are made to accommodate real-world city-scale data. It includes over 100 power network nodes and more than 200 heat network nodes. Both wind and PV generation are included, with their baseline power profiles randomly sampled from historical data. A hurricane makes landfall from the southeast and moves northward. During the hazard period, PV output is set to zero. When wind speed exceeds the cut-out threshold of 25 m/s, wind turbines are assumed to be disconnected from the grid, and wind power output is set to zero.

Further details of the systems and hazard evolution can be found in \cite{Wang_case_data_2024}, including materials and files relevant to the numerical experiments in this study. Simulations are executed on a computer equipped with an Intel i9-12900H CPU, using Python 3.12 and Gurobipy to solve the optimization problems.

\subsection{Extreme weather risk assessment and multi-type emergency decision-making}\label{risk_assessment}

First, based on system\#1, the full process and results of the proposed framework are reported. Monte Carlo simulation is used to compute the expected value and error bounds of the systems' objectives. The performance metrics of the networks are calculated by dividing the objective function values obtained from simulations by the baseline values under disaster-free conditions, resulting in a numerical range between 0 and 1.

\begin{table}[!t]
	\centering
	\setlength{\tabcolsep}{0.5mm}
	\renewcommand\arraystretch{0.8}
	\caption{Case setup for numerical experiments.}
	\begin{tabular}{@{}lccccc@{}}
		\toprule
		 &
		  \multicolumn{1}{l}{\begin{tabular}[c]{@{}l@{}}{\textbf{Repair}}\\ {\textbf{Crew}}\end{tabular}} &
		  \multicolumn{1}{l}{\begin{tabular}[c]{@{}l@{}}{\textbf{Preventive}}\\ {\textbf{Reinforce}}\end{tabular}} &
		  \multicolumn{1}{l}{\begin{tabular}[c]{@{}l@{}}{\textbf{Topology}}\\ {\textbf{Reconfig}}\end{tabular}} &
		  \multicolumn{1}{l}{\begin{tabular}[c]{@{}l@{}}{\textbf{EV}}\\ {\textbf{Power Supply}}\end{tabular}} &
		  \multicolumn{1}{l}{\begin{tabular}[c]{@{}l@{}}{\textbf{Fuel}}\\ {\textbf{Transport}}\end{tabular}} \\ \midrule
		{Case\#1}    & \ding{55}    & \ding{55}  & \ding{55}  & \ding{55}  & ——         \\
		{Case\#2}    & Ideal Repair & \ding{55}  & \ding{55}  & \ding{55}  & ——         \\
		{Case\#3}    & \checkmark   & \ding{55}  & \ding{55}  & \ding{55}  & ——         \\
		{Case\#4}    & \checkmark   & \checkmark & \ding{55}  & \ding{55}  & ——         \\
		{Case\#5}    & \checkmark   & \ding{55}  & \checkmark & \ding{55}  & ——         \\
		{Case\#6}    & \checkmark   & \ding{55}  & \ding{55}  & \checkmark & ——         \\
		{Case\#7(*)} & \checkmark   & \ding{55}  & \ding{55}  & \ding{55}  & \ding{55}  \\
		{Case\#8(*)} & \checkmark   & \ding{55}  & \ding{55}  & \ding{55}  & \checkmark \\ \midrule
		\multicolumn{6}{l}{(*): Insufficient initial fuel reserves.}                   
	\end{tabular}
	\label{table-case-setup}
\end{table}

\normalsize

As shown in Table \ref{table-case-setup}, this paper sets up eight cases to demonstrate the scalability of the proposed framework, covering various emergency factors in an extreme weather event. Among these, preventive reinforcement selectively targeted 5 power lines and 10 transportation roads, reducing their damage probabilities or performance losses to 10\% of the original values. Meanwhile, ``Ideal Repair'' denotes repair operations that disregard the impact of rainstorms on the transportation network and resource constraints. For each case, at least 200 sampled scenarios are used in the Monte Carlo simulation to obtain the results.

\begin{figure}[!t]
	\centering
	\includegraphics[scale=0.335]{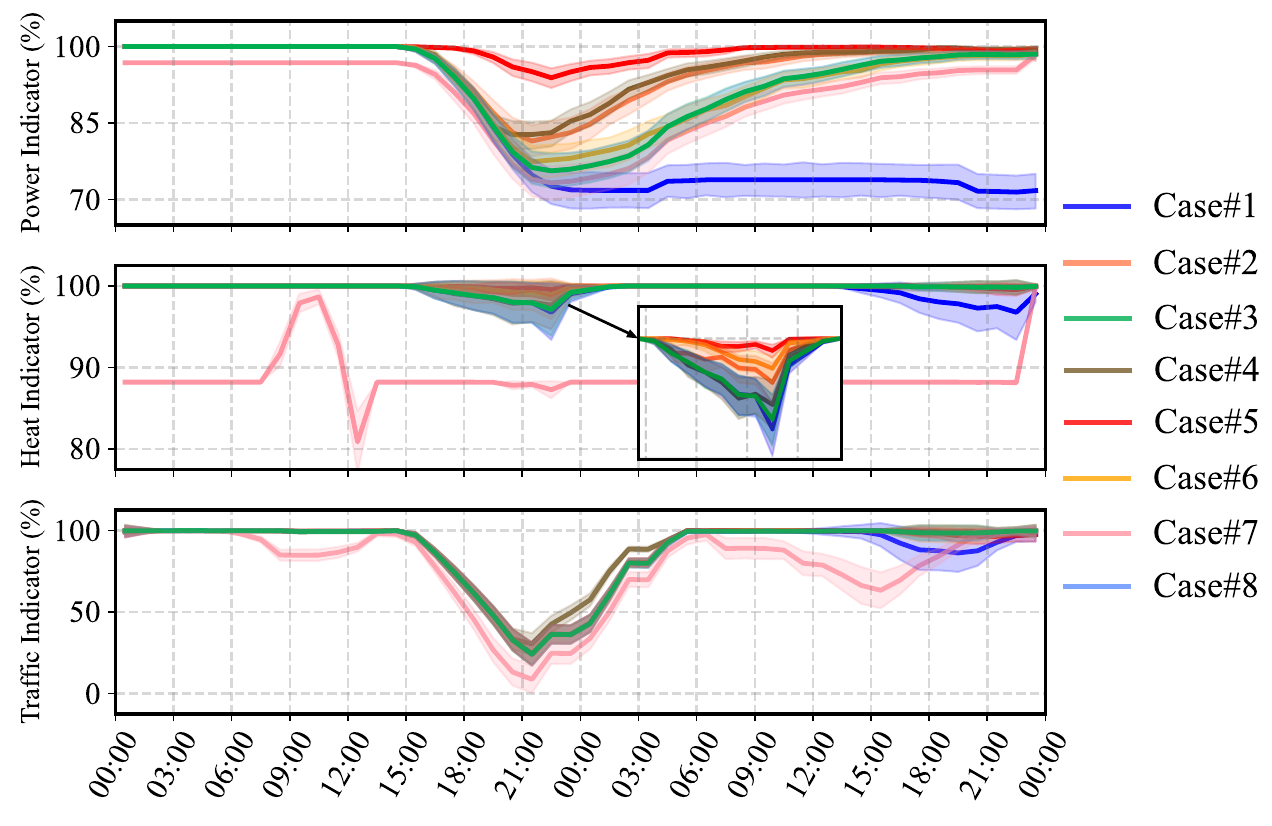}
	\caption{System\#1 indicator curves with error bounds for all cases.}
	\label{fig-compare-indicator-allcases}
\end{figure}

Figure \ref{fig-compare-indicator-allcases} illustrates that under different combinations of emergency decision factors, the power, heat, and transportation networks exhibit distinct performance curves. The solid lines denote the mean values, while the shaded areas indicate the corresponding confidence intervals. The narrow intervals show that the sample means provide an accurate approximation of the expected values of the underlying random variables.

\begin{figure}[!t]
	\centering
	\includegraphics[scale=0.330]{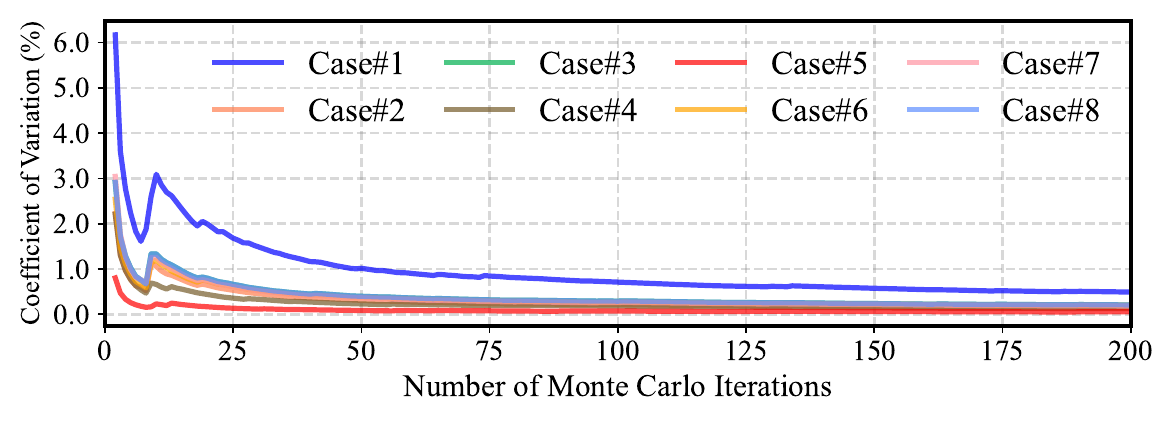}
	\caption{Coefficient of variation of system\#1's total objective function over Monte Carlo iterations.}
	\label{fig-CV-system1}
\end{figure}

\begin{figure}[!t]
	\centering
	\includegraphics[scale=0.340]{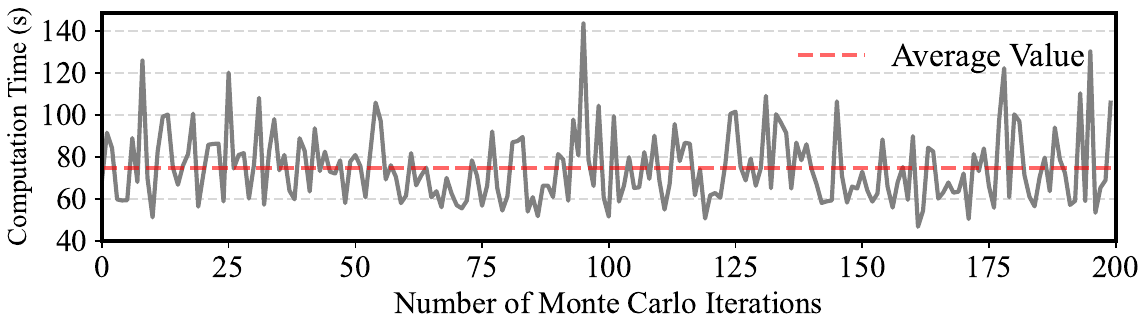}
	\caption{Computation time of each sample for system\#1 (case\#3) over Monte Carlo iterations.}
	\label{fig-Computation-Time-Per-Sample-system1-Case3}
\end{figure}

Convergence of the Monte Carlo simulation is evaluated using the coefficient of variation (CV). Using the total objective function in equation (\ref{Model-Complete-Form}) as the variable, the evolution of its CV is shown in Fig. \ref{fig-CV-system1}. As the simulation progresses, the CV decreases steadily, meeting the main requirements for CV in the literature \cite{Wang_Resilience_Constrained_UC_2018}. Figure \ref{fig-Computation-Time-Per-Sample-system1-Case3} uses case\#3 as an example and reports the computation time for each sample during the Monte Carlo simulation.

Furthermore, by incorporating multiple types of emergency decisions, various insights can be provided through combination and comparison.

Due to the destruction of power lines by the hurricane and the obstruction of traffic by the rainstorm, both the power and transportation systems suffer inevitable performance losses. Generally, the heat network is least affected as it is not directly attacked by the hurricane and rainstorm, but it still experiences performance degradation due to interdependencies. Without any emergency decisions (Case\#1), the power system cannot recover to its normal state after being hit by the disaster. The heat network can maintain normal performance during off-peak hours, and the transportation network's performance can gradually recover as road-accumulated water subsides.

Each added emergency decision enhances the ability of the energy-transportation coupled system to maintain performance during extreme weather events. Power system repairs (Case\#3) allow for gradual restoration of power supply and performance. Compared to ideal repairs (Case\#2), actual repairs are constrained by the transportation network's performance and resource availability, with delays caused by rain-induced waterlogging postponing the repair of damaged lines, thus slowing the recovery speed. Preventive reinforcement (Case\#4) directly improves each system's performance during the disaster, especially since it reinforced not only the power lines but also the transportation roads to ensure critical mobility. Topology reconfiguration (Case\#5) significantly enhances the service performance of the power and heat networks by avoiding interruptions to certain critical power and heat loads. EV power supply (Case\#6), through discharging, temporarily assists in supplying certain critical power loads and indirectly helps protect heat loads.


If the initial fuel reserve is insufficient (Case\#7), its impact will ripple through all systems. Power and heat sources will be constrained by fuel shortages, forcing reduced output and load shedding. The transportation system will suffer performance losses due to charging limitations. However, 
even with insufficient fuel reserves, vehicle dispatch for fuel transport (Case\#8) can alleviate fuel shortages, promoting power and heat generation and pulling the system's performance curve back to that of fuel-sufficient conditions, closely overlapping with Case\#3.

\begin{figure}[!t]
	\centering
	\includegraphics[scale=0.303]{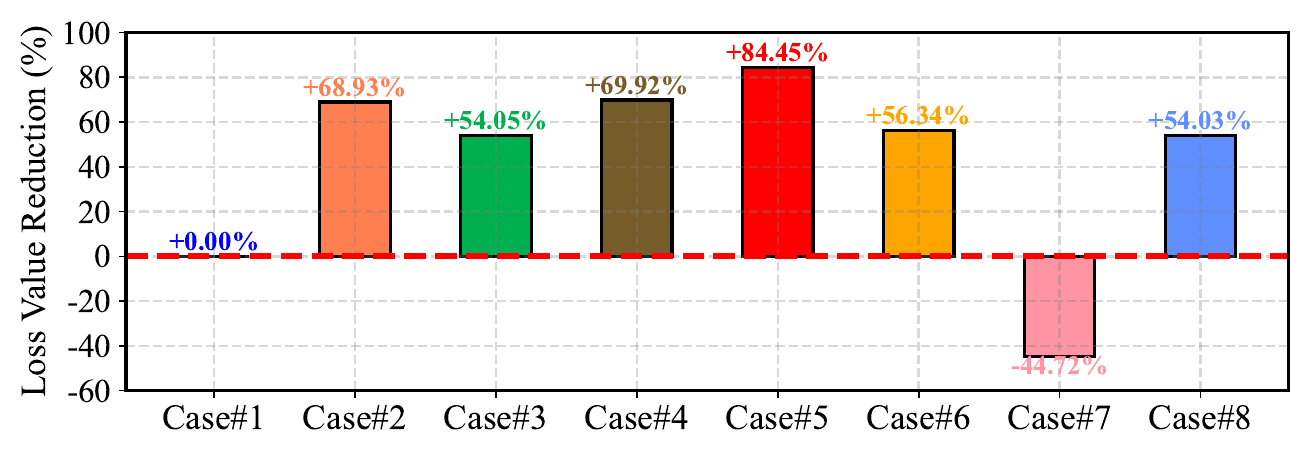}
	\caption{Contribution of different emergency decisions in reducing extreme weather losses for system\#1.}
	\label{fig-emergency-decision-utility}
\end{figure}

By comparing with the baseline (Case\#1), the utility of each emergency decision can be quantified. As shown in Fig. \ref{fig-emergency-decision-utility}, among single emergency decisions, topology reconfiguration (Case\#5) can reduce 84.45\% of the value loss caused by extreme weather, making it the most effective emergency decision in this context.

The above combinations and comparisons powerfully demonstrate the risk superposition and mutual support brought by energy-transportation coupling while also highlighting the necessity of comprehensively considering interdependencies among systems and multi-type emergency decisions.

\begin{figure}[!t]
	\centering
	\subfloat[]
	  {\includegraphics[scale=0.149]{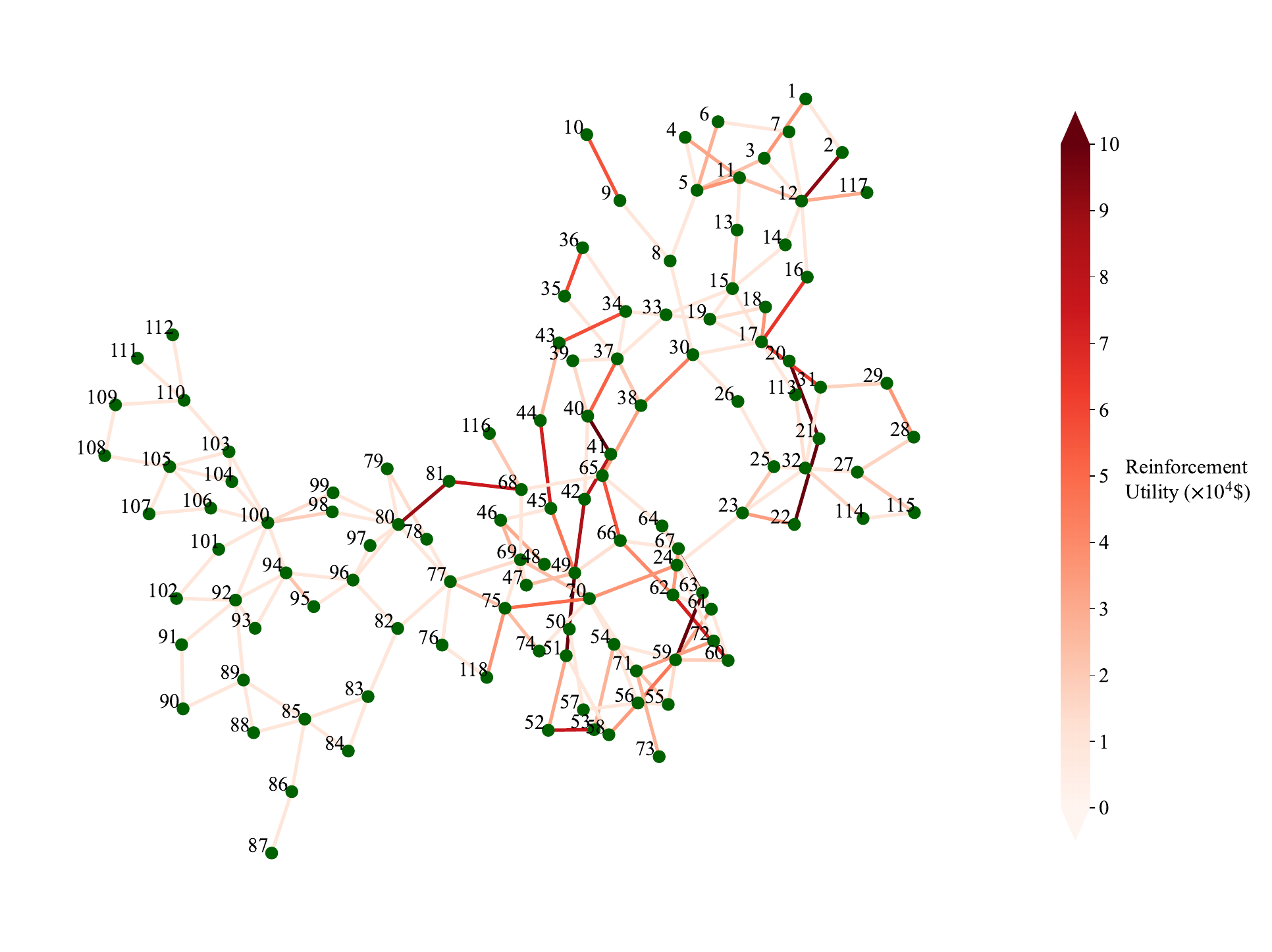}\label{fig-weakness-system2}}
	\hspace{0.0in}
	\subfloat[]
	  {\includegraphics[scale=0.124]{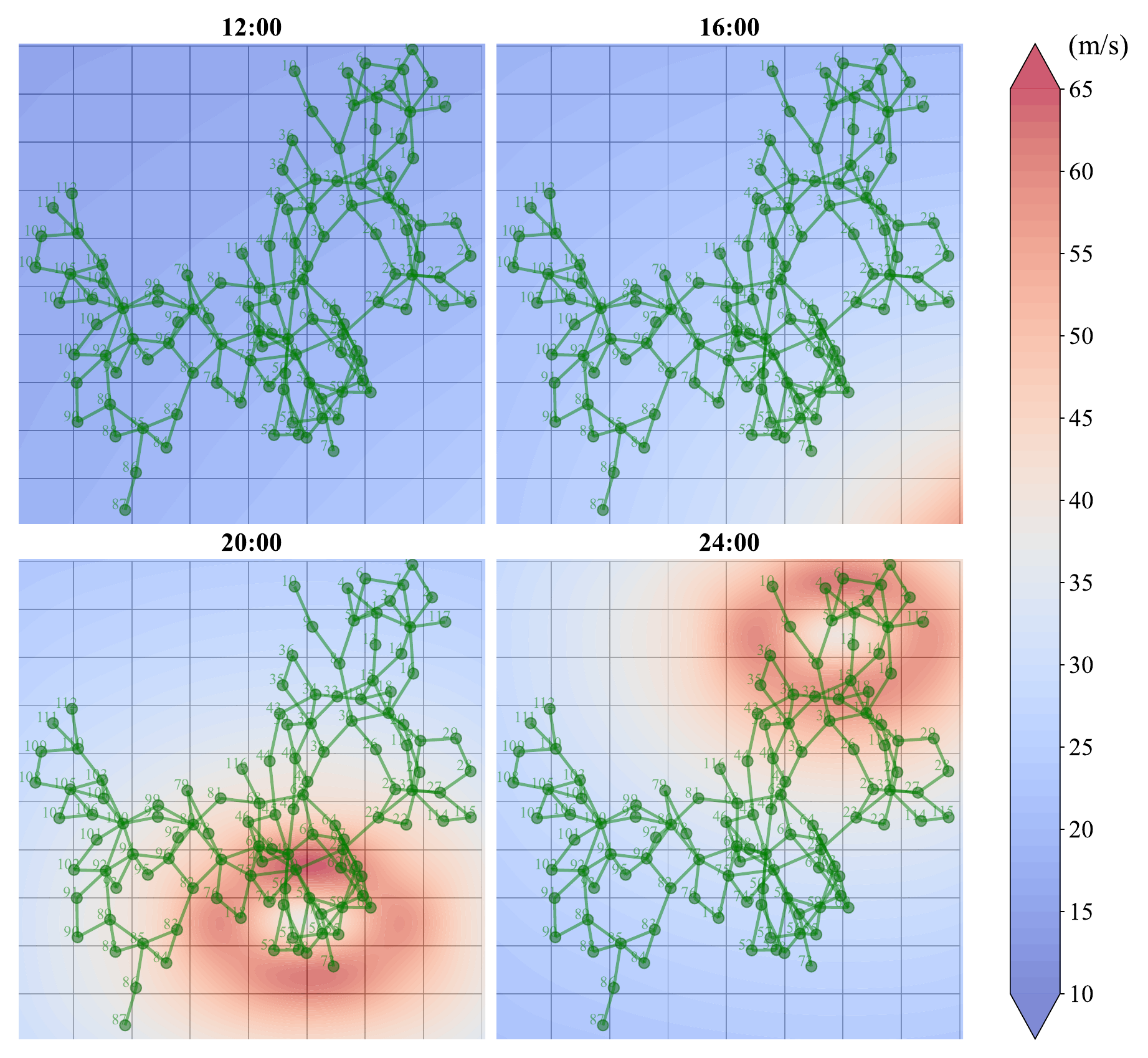}\label{fig-weather-time-space-system2}}
	\caption{Power network in system\#2. (a) Vulnerability indicator evaluated by simulation. (b) Wind speed distribution of hurricane disasters at certain moments.}
	\label{fig-weakness-weather-view-system2}
\end{figure}

Applying the proposed framework to system\#2 yields similar trends and conclusions. System\#2 is larger in scale and geographic coverage and faces a more severe hurricane, resulting in greater damage and requiring more resources and working hours for repairs. As illustrated in Fig. \ref{fig-weakness-weather-view-system2}, the vulnerable power lines of system\#2 are concentrated along the spatiotemporal trajectory of extreme weather events, highlighting the significance of geographical level in this study.

\subsection{Identification of system vulnerabilities}

For the task of identifying vulnerable power lines under extreme weather, the direct calculation of the indicator in Section \ref{Direct Assessment} is used as the benchmark. It reinforces each power line individually, runs full Monte Carlo simulations on the reinforced system, and quantifies the benefit of each reinforcement. This approach provides the most complete and accurate assessment, though it is also the most time-consuming, making it suitable as a reference standard.

In addition, the following methods are set up for comparison to highlight the advantages of the proposed framework combining models and surrogates.

\begin{itemize} 
	\item \textbf{Power Model Only (PMO)}: This method evaluates line vulnerability by considering only the power network, ignoring interdependencies with other systems. 
	
	\item \textbf{Surrogate Model Only (SMO)}: This reflects efforts to address the task using data-driven methods \cite{Cassottana_Predicting_Resilience_2022}.
	Using data from Monte Carlo simulations, a neural network surrogate for the coupled energy-transportation system is trained. The input consists of extreme weather scenarios, and the output is the system performance indicator. This method directly evaluates line vulnerability via the surrogate. 
	
	\item \textbf{Heuristic Loss Allocation (HLA)}: This draws inspiration from the work represented by \cite{LIAN2023_cascading_failure_graph}.
	This method heuristically attributes losses from $t-1$ to $t$ to the power line that was in a failed state during this period. The losses are distributed across the lines proportionally, and the lines with the highest cumulative loss responsibility are deemed most critical for reinforcement. 
	
	\item \textbf{Hybrid Surrogate Model (HSM) proposed in this paper}: The heat and transportation networks are represented by neural network surrogates, combined with the power network, which uses a physical model. 
\end{itemize}

The vulnerability identification only requires capturing the underlying trends of interdependence, so a balance can be struck between accuracy and the complexity of the surrogate models. In the hybrid surrogate model (HSM) method, both the heat-network surrogate and the traffic-network surrogate use 2,048 generated samples each, split into training and test sets at an 8:2 ratio. Both neural networks use two fully connected layers with ReLU activation and the Adam optimizer. After hyperparameter tuning, the heat-network surrogate uses a learning rate of 0.0002, a batch size of 64, and an L2 regularization coefficient of 0.016. The traffic-network surrogate uses a learning rate of 0.0008, a batch size of 512, and an L2 coefficient of 0.038. Training runs for up to 10,000 epochs with early stopping, using 20\% of the training set as a validation set and stopping when no improvement appears over consecutive validation checks. After training these regression models, performance is evaluated on the test set using the mean absolute percentage error (MAPE). Taking system\#1 as an example, the heat-network surrogate achieves a test-set MAPE of 0.075\%, and the traffic-network surrogate achieves 1.297\%.

\begin{figure}[!t]
	\centering
	\subfloat[]{\includegraphics[scale=0.375]{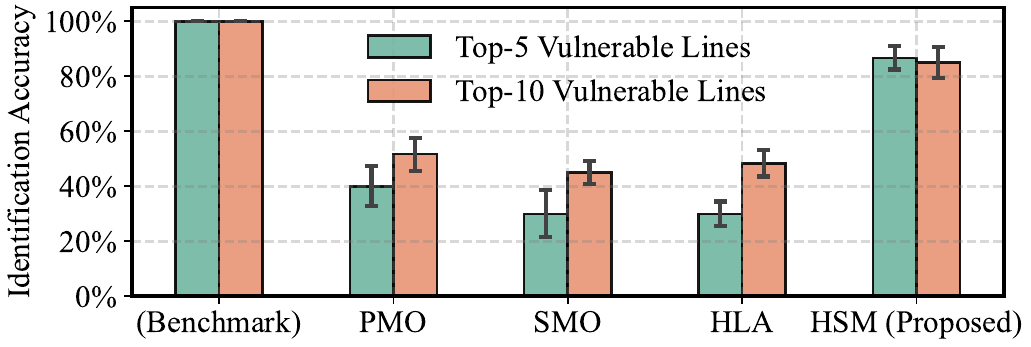}%
	\label{fig-repeat-system1}}
	\hspace{0.0in}
	\subfloat[]{\includegraphics[scale=0.375]{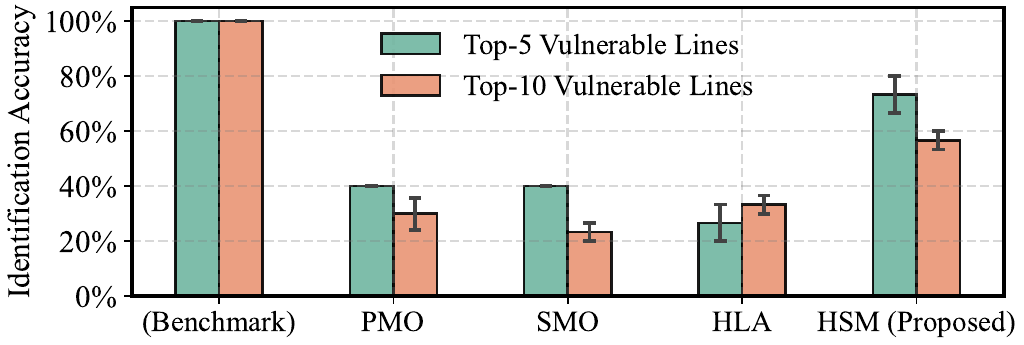}%
	\label{fig-repeat-system2}}
	\caption{Accuracy of different methods in identifying vulnerable lines. (a) System\#1. (b) System\#2.}
	\label{fig-repeat-experiments-accuracy}
\end{figure}

After all preparatory work, the same set of randomly sampled scenarios is used to evaluate each method. Additional details can be found in \cite{Wang_case_data_2024}. For system\#1 and system\#2, three repeat experiments were conducted for each with different random seeds. Figure \ref{fig-repeat-experiments-accuracy} shows the accuracy of each method in identifying the top 5 or 10 weakest lines.

On both system\#1 and system\#2, the proposed method outperforms all other baseline approaches, achieving the highest accuracy apart from the benchmark.

\begin{figure}[!t]
	\centering
	\subfloat[]
	  {\includegraphics[scale=0.375]{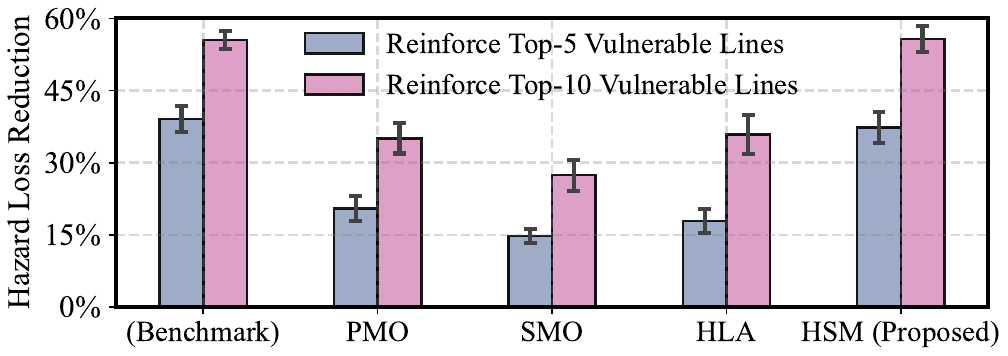}\label{fig-stat_stage2-system1}}
	\hspace{0.0in}
	\subfloat[]
	  {\includegraphics[scale=0.375]{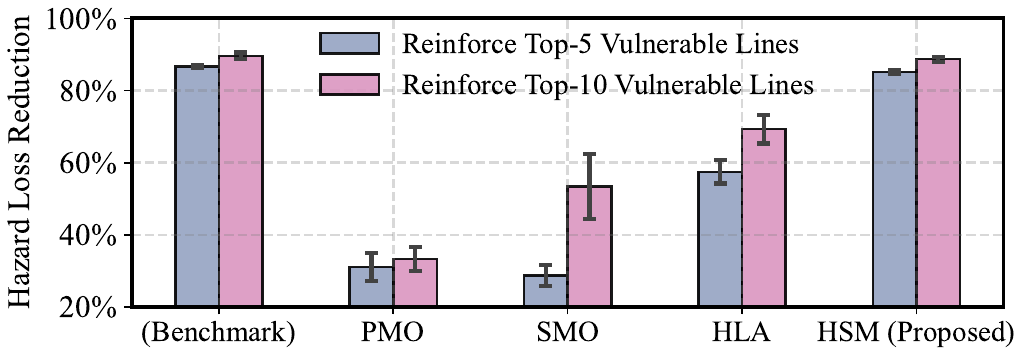}\label{fig-stat_stage2-system2}}
	\caption{Hazard loss reduction after reinforcing identified vulnerable lines. (a) System\#1. (b) System\#2.}
	\label{fig-harden-effect-show}
\end{figure}

Reinforcing the power lines identified by different methods significantly reduces system losses under the same extreme weather events. As shown in Fig. \ref{fig-harden-effect-show}, this highlights the value of assessing system vulnerabilities. While HSM does not achieve 100\% accuracy, most errors arise from non-critical lines. Reinforcing the top-ranked lines identified by HSM yields nearly the same loss reduction as the benchmark, indicating that the proposed method successfully captures key system patterns and identifies the most critical lines. In contrast, other baseline methods suffer from lower accuracy, leading to less effective reinforcement and weaker risk mitigation. This suggests that they fall short in capturing the overall risk trends and fail to provide reliable guidance for reinforcement decisions.

Among the baseline methods, PMO can be regarded as the most direct ablation study of the proposed HSM. Table \ref{table-computational-efficiency-show} reports the optimization problem sizes, from small to large systems, together with the computational performance of completing the vulnerability identification, to demonstrate scalability. Compared to the full direct assessment benchmark, HSM significantly reduces the number of constraints and variables, only slightly above that of PMO, achieving significant acceleration. At the same time, HSM retains consideration of interdependencies across systems, which is essential for evaluating coupled systems.

\begin{table}[!t]
	\centering
	\setlength{\tabcolsep}{1.0mm}
	\renewcommand\arraystretch{0.8}
	\caption{Comparison of problem scale and computational performance between HSM and PMO.}
	\begin{tabular}{@{}lccc@{}}
	\toprule
	\multicolumn{1}{r}{}                                                            & {\textbf{Benchmark}}      & {\textbf{PMO}} & {\textbf{HSM (Proposed)}} \\ \midrule
	{\textbf{System\#1}}                                                            &                           &                &                           \\
	Variable Number                                                                 & $4.0\times 10^5$ (100\%)  & \ 6.08\%       & \ 9.01\%                  \\
	Constraint Number                                                               & $3.7 \times 10^5$ (100\%) & \ 8.84\%       & 12.21\%                   \\
	\begin{tabular}[c]{@{}l@{}}Vulnerability Identification\\ Time (s)\end{tabular} & $7.0\times 10^4$ (100\%)  & \ 1.03\%       & \ 3.75\%                  \\ \midrule
	{\textbf{System\#2}}                                                            &                           &                &                           \\
	Variable Number                                                                 & $7.0\times 10^5$ (100\%)  & 13.75\%        & 18.55\%                   \\
	Constraint Number                                                               & $7.1\times 10^5$ (100\%)  & 15.36\%        & 20.11\%                   \\
	\begin{tabular}[c]{@{}l@{}}Vulnerability Identification\\ Time (s)\end{tabular} & $9.9\times 10^4$ (100\%)  & 25.63\%        & 35.72\%                   \\ \bottomrule
	\end{tabular}
	\label{table-computational-efficiency-show}
\end{table}

\begin{figure}[!t]
	\centering
	\subfloat[]{\includegraphics[scale=0.335]{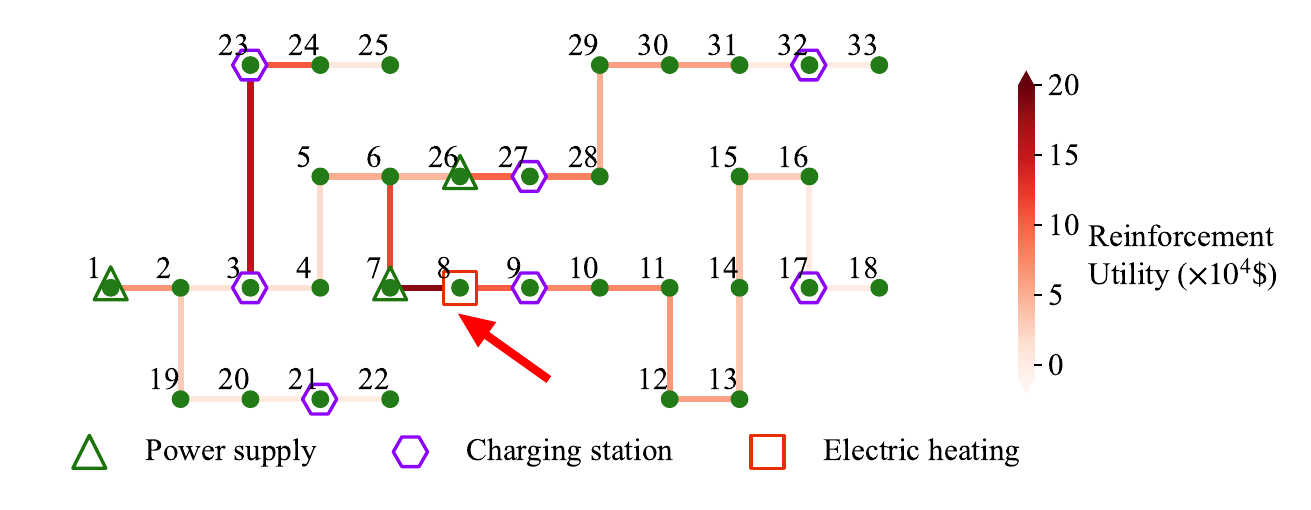}%
	\label{fig-weakness-DM}}
	\hfil
	\subfloat[]{\includegraphics[scale=0.335]{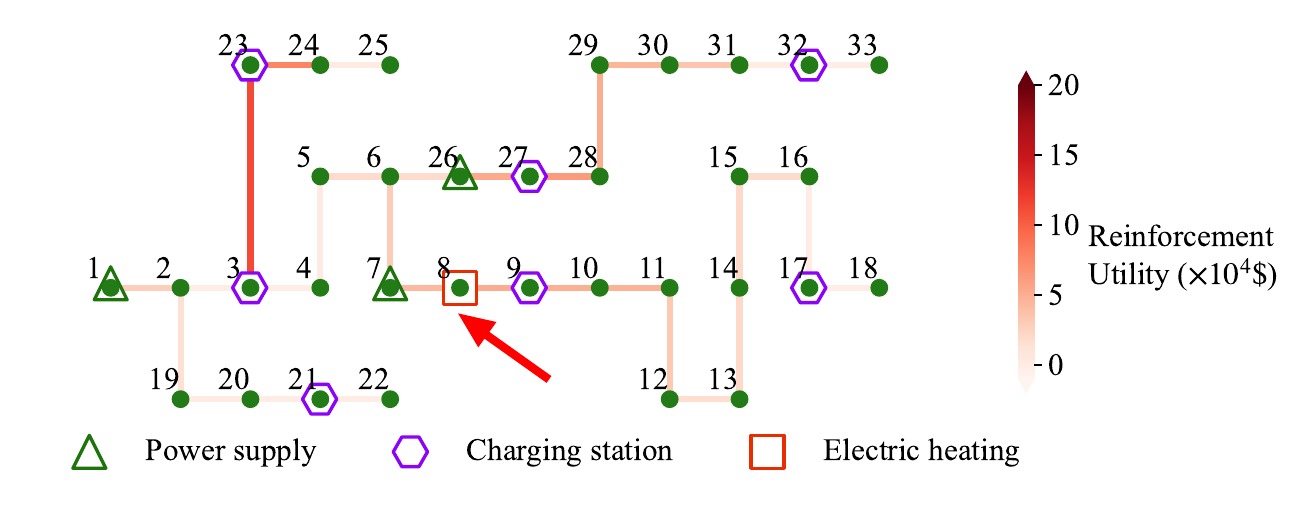}%
	\label{fig-weakness-OP}}
	\caption{Vulnerability indicator of power lines in system\#1. (a) Benchmark. (b) PMO.}
	\label{fig-compare-DM-OP}
\end{figure}

When evaluating with only the power network model (PMO), the identified vulnerabilities are limited to within the power system itself, missing interdependencies across systems. For instance, as shown in Fig. \ref{fig-compare-DM-OP}, line $(\#7-\#8)$ is the most critical in system\#1 due to electric heating at bus \#8. Failure of line $(\#7-\#8)$ not only disrupts electric loads but also significantly affects heat loads, causing substantial losses. The PMO method cannot capture this relationship and thus fails to identify this critical line.

The proposed HSM method strikes a balance among the other baseline approaches. Compared to the benchmark, it significantly reduces the problem size and accelerates assessment while preserving interdependencies and privacy. Compared to PMO, it partially incorporates interdependencies and captures essential coupling trends. Compared to HLA, it adopts a direct assessment framework, retaining the physical system's operational details. Compared to SMO, it employs a hybrid approach, using models and surrogates, rather than relying only on a purely data-driven approach to capture highly complex features and interdependencies. Moreover, SMO can only use the samples generated by the centralized risk-assessment procedure in Section \ref{risk_assessment}, and it is difficult for SMO to centrally integrate all subsystems to generate additional samples in advance. In contrast, in HSM, the heat-network and traffic-network surrogates are trained within themselves, enabling them to independently generate a large number of boundary-variable inputs covering a wide range as training samples in advance. This process does not require interaction with other systems.

In summary, the proposed HSM captures the system's main trends while achieving a balance between accuracy and speed.

\section{Conclusion}\label{Conclusion}

Under the threat of extreme weather events, urban infrastructure systems, particularly energy and transportation, face challenges of cross-system failure propagation and risk amplification. There are interdependencies and mutual support between systems, accompanied by multi-type energy-transportation emergency decisions. 
This paper proposes a framework to simulate and quantify the impacts of extreme weather on energy-transportation coupled systems. Using a unified network flow formulation, the framework integrates the heterogeneous energy and transportation sides, enabling full accommodation of multiple types of energy-transportation emergency decisions and capturing the compound spatiotemporal impacts of extreme weather on both systems simultaneously. Building on this framework, a method for identifying system vulnerabilities is further developed, employing neural network surrogates and their exact linearized representations to accelerate identification and preserve privacy.

The scenario-based approach used in this paper is constrained by scenario efficiency, and future work may explore simulation methods that improve Monte Carlo efficiency. In addition, the neural network surrogate models used for vulnerability identification may lack interpretability. Further research can focus on integrating causal inference mechanisms into the framework to enhance interpretability and guide targeted resilience improvements. These directions will be key areas for future research.




\bibliographystyle{ieeetr}
\bibliography{refs_abrv}

 




\vfill

\end{document}